\documentstyle[epsfig,12pt]{article}

\include{feynman}
\mark{{}{}}
\def\permil{\%\raise.10ex\hbox{$_{\scriptstyle 0}$}}

\begin{document}

%\begin{titlepage}

%\newpage
%\setcounter{page}{1}
\def\la{\mathrel{\mathpalette\fun <}} \def\ga{\mathrel{\mathpalette\fun >}}
\def\fun#1#2{\lower3.6pt%
\vbox{\baselineskip0pt\lineskip.9pt
  \ialign{$\mathsurround=0pt#1\hfil##\hfil$\crcr#2\crcr\sim\crcr}}}
%less than approximately and greater than approximately
\begin{titlepage}
\centerline{\LARGE Next-to-Leading Corrections to the BFKL Equation}
\vskip 0.4cm
\centerline{\LARGE  From the Gluon and Quark Production}
\vskip 0.4cm
\centerline{V. S. Fadin}
\vskip 0.3cm
\begin{center}
{
\it Budker Nuclear Physics Institute and Novosibirsk State University,
 630090, Novosibirsk, Russia}
\end{center}
\vskip 0.4cm
\centerline{L. N. Lipatov $^{\dagger}$}
\vskip 0.3cm

\begin{center}
{
\it Petersburg Nuclear Physics Institute,
Gatchina, 188350, St. Petersburg,  Russia }
\end{center}
%\vskip 0.3cm
\begin{center}
{
\it and}
\end{center}
%\vskip 0.3cm
\begin{center}
{
\it Deutsches Elektronen-Synchrotron DESY, Hamburg}
\end{center}
\vskip 15.0pt

\centerline{\bf Abstract}
\noindent
The gluon and quark production in the quasi-multi-Regge kinematics for
the final particles is considered. The differential cross-section for
different helicity states is calculated. The dimensional regularization
is used to remove the infrared divergencies in the corresponding
contributions to the BFKL equation. The other next-to-leading corrections
are discussed.
\vskip 3cm
\hrule
\vskip.3cm
\noindent

\noindent
$^{\dagger}${\it Humboldt Preistr\"ager\\
Work supported partly by INTAS and
the Russian Fund of Fundamental Investigations}
\vfill
\end{titlepage}
%\eject
%\textheight 210mm
%\topmargin 2mm
%\baselineskip=24pt

%\documentstyle[12pt,qqaajart]{article}

%\include{feynman}
%\setlength\topmargin{-0.5cm}
%\setlength\textwidth{16cm}
%\setlength\oddsidemargin{-0.1cm}
%\setlength\evensidemargin{-0.1cm}
%\headsep 30pt
%\mark{{}}
%\def\permil{\%\raise.10ex\hbox{$_{\scriptstyle 0}$}}
%\def\nll{ \nonumber \\}
%\input tcilatex

%\begin{document}

%\begin{titlepage}
%\author{V. S. Fadin \\
%Budker Nuclear Physics Institute, Novosibirsk, Russia \\
%L. N. Lipatov $^{\dagger}$\\
%Petersburg Nuclear Physics Institute, Gatchina, Russia \\}
%\title{Production of gluon and quark pairs with a fixed invariant mass \\
%at high energy \\}
%\vskip 15.0pt
%\centerline{\bf Abstract}
%\noindent
%Next to leading corrections to the BFKL equation related with the gluon and
%quark pair production in the multi-Regge kinematics are calculated.
%\vskip 3cm
%\noindent
%\noindent
%$^ (\dagger )$ {\it Humboldt Preistr\"ager \\
%Work is supported partly by INTAS and the Russian Fund of Basic
%Reseaches}
%\vfill
%\end{titlepage}

\section{Introduction}

Now both GLAP [1] and BFKL [2] equations are used to describe the parton
distributions $n_i(x)$ at the small Bjorken variable $x$. The next to
leading corrections to the GLAP equation are well known. On the other hand,
the program of calculating the next-to-leading corrections to the BFKL
equation was formulated comparatively recently [3]. In this paper we
consider these corrections from the real gluon and quark production.

In each order of the perturbation theory the main contribution to the total
cross-section $\sigma _{tot}$ for the high energy collision of particles
with momenta $p_A$ and $p_B$ results from the multi-Regge kinematics for the
final state gluon momenta $k_0=p_{A^{\prime
}},k_1,...,k_n,k_{n+1}=p_{B^{\prime }}$ (see Fig.1):

\begin{eqnarray}
s \gg s_{i}=2k_{i-1}k_i \gg t_i=q_i^2\,,\,q_i=p_A-\sum\limits_{r=0}^{i-1}
k_r\,,\,\,
\prod\limits_{i=1}^{n+1} s_i=s\prod\limits_{i=1}^n {\bf k}_{i}^2 \,\,,\,\,
k_{\perp}^2=-{\bf k}^2\,,
\end{eqnarray}
where $k_{i\perp}$ are transverse components of momenta $k_i$.

In the leading logarithmic approximation (LLA) the $n$-gluon production
amplitude in this kinematics has the multi-Regge form [2]:
\begin{eqnarray}
A_{2+n}^{LLA}=A_{2+n}^{tree} \prod\limits_{i=1}^{n+1}
s_i^{\omega(t_i)}\,\,,
\end{eqnarray}
where $j=1+\omega (t)$ is the gluon Regge trajectory and
\begin{eqnarray}
\omega(t)=-\frac{g^2 N_c}{16\pi^3}\int d^2{\bf k}
\frac{{\bf q}^2}{{\bf k}^2 ({\bf q-k})^2} \,\,\,,\,\,t=-{\bf q}^2 \,\, .
\end{eqnarray}

The infrared divergencies in the Regge factors $s^{\omega (t_i)}$ cancel in
the total cross section with contributions from the real gluons. The
production amplitude for the gluon-gluon scattering in the tree
approximation has the following factorized form

\begin{eqnarray}
A_{2+n}^{tree}=2 s gT_{a'a}^{c_1}\Gamma_{A'A}\frac{1}{t_1}gT_{c_2 c_1}^{d_1}
\Gamma_{2,1}^1\frac{1}{t_2}....
gT_{c_{n+1}c_n}^{d_n}\Gamma_{n+1,n}^n
\frac{1}{t_{n+1}}gT_{b'b}^{c_{n+1}}\Gamma_{B'B}\,\,.
\end{eqnarray}
Here $a,b$ and $a^{\prime},b^{\prime},d_r $ ($r=1,2...n$) are colour indices
for initial and final gluons correspondingly. $T_{ab}^{c}=-if_{abc}$ are
generators of the gauge group $SU(N_c)$ in the adjoint representation, $g$
is the Yang-Mills coupling constant,

\begin{equation}
\Gamma _{A^{\prime }A}=\delta _{\lambda _A,\lambda _A^{\prime
}}\,\,,\,\,\,\Gamma _{r+1,r}^r=C_\mu (q_{r+1},q_r)\overline{e}_\mu (k_r)
\end{equation}
are the reggeon-particle-particle (RPP) and reggeon-reggeon-particle (RRP)
vertices correspondingly; $e(k_r)$ is the polarization vector of the
produced gluon. The quantity $\lambda _r=\pm 1$ is the helicity of the gluon
$r$ in the c.m.system. In LLA the $s$-channel helicities of colliding
particles are conserved. The effective nonlocal RRP vertex $C(q_2,q_1)$ is
given below [2]
\begin{eqnarray}
C(q_2,q_1) = -q_1-q_2 +
p_A(\frac{q_1^2}{k_1p_A}+2\frac{k_1p_B}{p_Ap_B}) -
p_B(\frac{q_2^2}{k_1p_B}+2\frac{k_1p_A}{p_Ap_B}).
\end{eqnarray}
It has the important property corresponding to the current conservation
\begin{eqnarray}
(k_1)_\mu C_\mu (q_2,q_1)=0,
\end{eqnarray}
which gives us a possibility to chose an arbitrary gauge for each of the
produced gluons. For example, in the left ($l$) light cone gauges, where $%
p_Ae^l(k)=0$ and $ke^l(k)=0$, one can use the following parametrisation of
the polarization vector $e^l(k)$
\begin{eqnarray}
e^l=e_{\perp}^l-\frac{k_{\perp}e_{\perp}^l}{kp_A} p_A
\end{eqnarray}
in terms of the two dimensional vector $e_{\perp }^l$. In this gauge the RRP
vertex takes an especially simple form, if we introduce the complex
components $e=\overline{e}_x+i\overline{e}_y\,,e^{*}=\overline{e}_x-i
\overline{e}_y$ and $k=k_x+ik_y\,,k^{*}=k_x-ik_y$ for transverse vectors $
\overline{e}_{\perp }^l,k_{\perp }$ [4]
\begin{eqnarray}
\Gamma_{2,1}^1=C e^* + C^* e ,\,\,\, C=\frac{q_1^* q_2}{k_1^*}.
\end{eqnarray}
The integral kernel of the BFKL equation in LLA is expressed in terms of the
product of the effective vertices (real contribution) and the gluon Regge
trajectory (virtual contribution). The above complex representation of the
effective vertex was used in ref. [4] to construct an effective scalar field
theory for the multi-Regge processes. It was derived recently from QCD by
integrating over the fields which correspond to the highly virtual particles
produced in the multi-Regge kinematics in the direct channels $s_i$ [5].

The total cross-section calculated in LLA using the above expressions for
production amplitudes grows very rapidly as $s^\omega $ ($\omega
=(g^2N_c/\pi ^2)\ln 2$) and violates the Froissart bound $\sigma _{tot}<c\ln
{}^2s$ [2]. One of the possible ways of the unitarization of the scattering
amplitudes corresponds to the solution of the BKP equations [6] for
multi-gluon compound states. These equations have a number of remarkable
properties, including conformal invariance [7], holomorphic separability
[8], and existence of nontrivial integrals of motion [9]. The Hamiltonian
for the corresponding Schr\"odinger equations coincides with the Hamiltonian
for a completely integrable Heisenberg model with the spins belonging to an
infinite dimensional representation of the noncompact M\"obius group [10].

All these results are based on calculations of effective Reggeon vertices
and the gluon Regge trajectory in the first nontrivial orders of
perturbation theory. Up to now we do not know the region of applicability of
LLA including the intervals of energies and momentum transfers fixing the
scale for the QCD coupling constant. The simple form of the BFKL equation
summing contributions of the Feynman diagrams describing the Pomeron as a
compound state of two reggeized gluons is valid also in the next-to-leading
approximation. To go beyond LLA the Born amplitudes for a quasi-multi-Regge
kinematics of produced gluons were calculated [3] and one-loop corrections
to the reggeon-reggeon-particle vertex were found [11,12]. Also two loop
contributions to the gluon Regge trajectory were calculated [13]. For the
total correction to the BFKL equation only the contribution related to the
production of a pair of gluons or quarks with a fixed invariant mass is not
known. This paper is devoted to the solution of this problem.

It is enough to consider the simplest process, in which in the final state
we have two gluons with momenta $p_{A^{\prime }}$ and $p_{B^{\prime }}$
almost coinciding with momenta $p_A$ and $p_B$ of initial gluons and a group
of particles in the central rapidity region (see Fig.2). The momenta $q_1$
and $q_2$ of virtual gluons in crossing channels $t_1$ and $t_2$ in this
kinematics can be decomposed as follows:%
\begin{eqnarray}
q_1=q_{1\perp}+\beta \,p_A\,,\,\,q_2=q_{2\perp}-\alpha \,p_B\,
\end{eqnarray}
where $\beta $ and $\alpha $ are the Sudakov parameters of the total
momentum $k=\sum\limits_{i=1}^nk_i$ for the produced particles:
\begin{eqnarray}
k=k_\perp + \beta \,p_A + \alpha \,p_B\,\, ,\,\, \kappa =k^2=s\alpha \beta +
(q_1-q_2)_\perp^2
\end{eqnarray}
and $\sqrt{\kappa }$ is their invariant mass which is asumed to be fixed at
high energies: $\kappa \ll s$.

The production amplitude in this kinematics has a factorised form and it is
expressed in terms of the scattering amplitude for two virtual gluons which
are reggeized after taking into account the radiative corrections. For its
gauge invariance in a general case of the multi gluon production one should
introduce an infinite number of the induced vertices for the interaction of
the reggeised gluon with usual gluons. The Feynman rules for these vertices
can be derived from the effective action [14] local in the particle
rapidities
\begin{eqnarray}
S_{eff}=-\int d^4 x \,\,tr \,\{\frac{1}{2}G_{\mu \nu}^2(v)+[A_-(v_-)-
A_-]\,\partial^2_\sigma A_++[A_+(v_+)-A_+]\,
\partial^2_\sigma A_-\}\,.
\end{eqnarray}
Here the fields $A_{\pm }$ and $v_\alpha $ are the anti-hermitial $SU(N_c)$%
-matrices describing the reggeised and usual gluons correspondingly; $v_{\pm
}$ are the light-cone components of $v_\alpha $. The quantities $A_{\pm
}(v_{\pm })$ are the composite fields expressed through the gluon fields $%
v_{\pm }$ entering in the Wilson $P$-ordered exponents:
\begin{equation}
A_{\pm }(v_{\pm })=-\frac 1g\partial _{\pm }P\,\exp (-\frac g2\int_{-\infty
}^{x^{\pm }}dx^{\prime \pm }v_{\pm })
\end{equation}
The reggeon fields obey to the additional constraint $\partial _{\mp }A_{\pm
}=0$ , which is important for the gauge invariance of the effective action.
We should add to the usual Yang-Mills action $\frac 12G_{\mu \nu }^2(v)$
also the term which describes the quark interactions. Similar to the pomeron
case [15] one can construct the Reggeon calculus in QCD for the Reggeon
fields $A_{\pm }$\thinspace starting from the above effective action and
integrating the functional integral over the gluon fields $v$ [14]. Because
$%
A_{\pm }(v_{\pm })$ has a linear term in $v_{\pm }$ the classical extremum
of $S_{eff}$ is situated at non-vanishing values of $v$ satisfying to
gauge-invariant Euler-Lagrange equations. Using the gaussian approximation
for the quantum fluctuations near this classical solution one could find
one-loop corrections to the BFKL kernel in an independent way in comparison
with the dispersion method of refs.[3],[11]. The possible advantage of this
approach could be a better infrared convergency of intermediate expressions.

In the next section we reproduce the gluon and quark production amplitudes
in the quasi-multi-Regge kinematics. In the third section the properties of
the amplitudes with the definite helicities of the final particles are
discussed. In the fourth section the real corrections to the BFKL equation
are constructed in terms of the integrals from bilinear combinations of
helicity amplitudes. In the fifth section the infrared divergencies in the
integrals over momenta of produced particles are extracted and regularized.
In Conclusion the obtained results are discussed.

\section{Gluon and quark pair production in the quasi-multi-Regge kinematics}

It is known (see [3],[14]) , that the gluon production amplitude $%
A_{2\rightarrow 4}$ in the quasi-multi-Regge kinematics for final particles
is expressed through the tensor $A^{\alpha _1\alpha _2}$ describing the
transition of two reggeized gluons with momenta $q_1,-q_{2}$ into two real
gluons with momenta $k_1,k_2$:

\begin{equation}
A_{2\rightarrow 4}=-g\,p_A^{+}\,T_{a^{\prime }a}^{c_1}\,
\delta_{\lambda_{A^{\prime}}\lambda_A} \,\frac 1{t_1}\,\psi
_{d_1d_2c_2c_1}^{\alpha _1\alpha _2}\,\frac 1{t_2}\,g\,p_B^{-}\,T_{b^{\prime
}b}^{c_2}\, \delta_{\lambda_{B^{\prime }}\lambda_B}\,,
\end{equation}

\begin{equation}
\psi _{d_1d_2c_2c_1}^{\alpha _1\alpha _2}=2\,g^2\left[
T_{d_1d}^{c_1}\,T_{d_2d}^{c_2}\,A^{\alpha _1\alpha
_2}(k_1,k_2)+T_{d_2d}^{c_1}\,T_{d_1d}^{c_2}\,A^{\alpha _2\alpha
_1}(k_2,k_1)\right] .
\end{equation}
Here $p_A^{+}=n_\alpha ^{+}p_A^\alpha \,=\sqrt{s}$ and $p_B^{-}=n_\alpha
^{-}p_B^\alpha =\sqrt{s}$ are the light-cone components of the colliding
particle momenta. The invariants $t_i$ are expressed through momentum
transfers: $t_i=q_i^2=q_{i\perp }^2$. $a,b$ and $a^{\prime },b^{\prime }$
are the colour indices of initial and scattered gluons correspondingly. The
helicity $\lambda _i$ of each colliding particle is conserved ($\lambda
_i=\lambda _{i^{\prime }}$). The matrices $T$ are the colour group
generators in the adjoint representation with the commutation relations: $%
\left[ T^k,T^l\right] =if^{klr}T^r$. The produced gluons with the Sudakov
momenta $k_i^{\pm },k_i^{\perp }$ ($i=1,2$) have the colour and Lorentz
indices $d_i$ and $\alpha _i$ correspondingly. The tensor $A^{\alpha
_1\alpha _2}(k_1,k_2)$ can be written as the sum of contributions of several
Feynman's diagrams (see Fig.3) [14]:%
$$
A^{\alpha _1\alpha _2}=-\frac{\Gamma ^{\alpha _1\beta
-}(k_1,k_1-q_1)\,\Gamma ^{\alpha _2\beta +}(k_2,k_2+q_2)}{2\,\,(q_1-k_1)^2}-
\frac{\gamma ^{\alpha _2\alpha _1\beta }(k_2,-k_1)\,\Gamma ^{+-\beta
}(q_2,q_1)}{2\,\,(k_1+k_2)^2}
$$
\begin{equation}
+n^{+\alpha _1}n^{-\alpha _2}-g^{\alpha _1\alpha _2}-\frac 12n^{+\alpha
_2}n^{-\alpha _1}+t_1\frac{n^{-\alpha _1}\,n^{-\alpha _2}}{%
k_1^{-}(k_1^{-}+k_2^{-})}+t_2\frac{n^{+\alpha _1}\,n^{+\alpha _2}}{%
k_2^{+}(k_1^{+}+k_2^{+})}\,,
\end{equation}
where we have $g^{00}=-g^{ii}=1$ for non-zero components of $g^{\alpha
_1\alpha _2}$and the light-cone vectors $n^{\pm }$ are $n^{+}=p_BE^{-1},%
\,n^{-}=p_AE^{-1},\,4E^2=s$\thinspace . The Yang-Mills vertex

\begin{equation}
\gamma ^{\alpha _2\alpha _1\beta }(k_2,-k_1)=(k_2-k_1)^\beta g^{\alpha
_2\alpha _1}+(2k_1+k_2)^{\alpha _2}g^{\alpha _1\beta }-(2k_2+k_1)^{\alpha
_1}g^{\alpha _2\beta }
\end{equation}
 enters also in the effective vertices $\left[ 2\right] $:

$$
\Gamma ^{\alpha _1\beta -}(k_1,k_1-q_1)=\gamma ^{\alpha _1\beta
-}(k_1,k_1-q_1)-t_1n^{-\alpha _1}\frac 1{k_1^{-}}n^{-\beta},
$$
$$
\Gamma^{\alpha_2 \beta +}(k_2,k_2+q_2)=\gamma^{\alpha_2 \beta
+}(k_2,k_2+q_2)- t_2n^{+\alpha}\frac{1}{k_2^+}n^{+\beta},
$$
\begin{equation}
\Gamma ^{+-\beta }(q_2,q_1)=\gamma ^{+-\beta }(q_2,q_1)-2t_1\frac{n^{-\beta
} }{q_1^{-}-q_2^{-}}+2t_2\frac{n^{+\beta }}{q_1^{+}-q_2^{+}}\,.
\end{equation}

Let us consider the gauge properties of $A^{\alpha_1\alpha_2}$. Taking into
account that particles $1$ and $2$ are on the mass shell and using the
Ward-Slavnov identity for the production vertex:
\begin{equation}
(k_2+k_1)^\beta \Gamma ^{+-\beta }(q_2,q_1)=0,\,\,\,q_1-q_2=k_1+k_2 ,
\end{equation}
we can perform in $k_1^{\alpha_1}A^{\alpha_1 \alpha_2}$ the substitution:

\begin{equation}
k_1^{\alpha _1}\gamma ^{\alpha _2\alpha _1\beta }(k_2,-k_1)\rightarrow
k_2^{\alpha _2}k_1^\beta -(k_1+k_2)^2g^{\alpha _2\beta }.
\end{equation}
With the use of the expression for the light cone projection of the
Yang-Mills vertex

\begin{equation}
\gamma ^{\alpha _1\beta -}(k_1,k_1-q_1)=(2k_1-q_1)^{-}g^{\alpha _1\beta
}+(2q_1-k_1)^{\alpha _1}n^{-\beta }-(q_1+k_1)^\beta n^{-\alpha _1},
\end{equation}
one can derive the Ward-Slavnov identity for the scattering vertex:

\begin{equation}
k^{\alpha _1}\Gamma ^{\alpha _1\beta -}(k_1,k_1-q_1)=-(q_1-k_1)^2n^{-\beta
}+(k_1-q_1)^\beta k_1^{-}.
\end{equation}

>From above relations we obtain the following gauge property of $A^{\alpha_1
\alpha_2}$:

$$
k_1^{\alpha _1}A^{\alpha _1\alpha _2}=\frac 12(\Gamma ^{\alpha
_2-+}(k_2,k_2+q_2)+\Gamma ^{+-\alpha _2}(q_2,q_1))
$$
$$
+\frac 12\left( k_1^{-}\frac{(k_2+q_2)^\beta \Gamma ^{\alpha _2\beta
+}(k_2,k_2+q_2)}{(k_2+q_2)^2}-k_2^{\alpha _2}\frac{k_1^\beta \Gamma
^{+-\beta }(q_2,q_1)}{(k_1+k_2)^2}\right)
$$
$$
+k_1^{+}n^{-\alpha _2}-\frac 12k_1^{-}n^{+\alpha _2}-k_1^{\alpha _2}+t_1
\frac{n^{-\alpha _2}}{k_1^{-}+k_2^{-}}+t_2\frac{k_1^{+}}{k_2^{+}}\frac{%
n^{+\alpha _2}}{k_1^{+}+k_2^{+}}=
$$
\begin{equation}
=\frac 12k_2^{\alpha _2}\left( \frac{k_1^{-}k_2^{+}}{(k_2+q_2)^2}-\frac{%
k_1^\beta \Gamma ^{+-\beta }(q_2,q_1)}{(k_1+k_2)^2}\right) .
\end{equation}
It means, that the production amplitude $A_{2\rightarrow 4}$ multiplied by
the gluon polarization vectors $e(k_i)$ is gauge invariant and the
contribution of the Faddeev-Popov ghosts to the final state is fixed.
However, instead of working with a covariant gauge we shall use the
light-cone gauges different for two produced gluons $\left[ 14\right] $:

\begin{equation}
e^\alpha (k_1)k_1^\alpha =e^{-}(k_1)=0\,,\,\,\,e^\alpha (k_2)k_2^\alpha
=e^{+}(k_2)=0.
\end{equation}
The polarization vectors can be parametrized as follows

\begin{equation}
e(k_1)=e_{1\perp }-\frac{(k_1e_{1\perp })}{k_1^{-}}n^{-},\,\,e(k_2)=e_{2%
\perp }-\frac{(k_2e_{2\perp })}{k_2^{+}}n^{+},
\end{equation}
corresponding to the polarization matrices

\begin{equation}
\Lambda ^{\alpha _1\alpha _1^{\prime }}(k_1)=-g^{\alpha _1\alpha _1^{\prime
}}+\frac{k_1^{\alpha _1}n^{-\alpha _1^{\prime }}+k_1^{\alpha _1^{\prime
}}n^{-\alpha _1}}{k_1^{-}}\,,
\end{equation}
\begin{equation}
\Lambda ^{\alpha _2\alpha _2^{\prime }}(k_2)=-g^{\alpha _2\alpha _2^{\prime
}}+\frac{k_2^{\alpha _2}n^{+\alpha _2^{\prime }}+k_2^{\alpha _2^{\prime
}}n^{+\alpha _2}}{k_2^{+}}\,.
\end{equation}

In these gauges for $e(k_i)$ the matrix element of the tensor $A^{\alpha
_1\alpha _2}$ can be expressed in terms of a new tensor $a^{\alpha _1\alpha
_2}$ with pure transverse components using the definition:

\begin{equation}
e_{\alpha _1}^{*}(k_1)\,e_{\alpha _2}^{*}(k_2)\,A^{\alpha _1\alpha _2}\equiv
e_{\alpha _1}^{\perp *}(k_1)\, e_{\alpha _2}^{\perp *}(k_2)\,a^{\alpha
_1\alpha _2}.
\end{equation}

To find this tensor we should perform the following substitutions in the
above expressions:
\begin{equation}
\Gamma ^{\alpha _1\beta -}(k_1,k_1-q_1)\rightarrow 2k_1^{-}g_{\perp
}^{\alpha _1\beta }+2(q_1^{\perp }-k_1^{\perp })^{\alpha _1}n^{-\beta },
\end{equation}
\begin{equation}
\Gamma ^{\alpha _2\beta +}(k_2,k_2+q_2)\rightarrow 2k_2^{+}g_{\perp
}^{\alpha _2\beta }-2(q_2^{\perp }+k_2^{\perp })^{\alpha _2}n^{+\beta };
\end{equation}
\begin{equation}
n^{+\alpha _1}n^{-\alpha _2}-g^{\alpha _1\alpha _2}-\frac 12n^{-\alpha
_1}n^{+\alpha _2}\rightarrow 2\frac{k_1^{\perp \alpha _1}k_2^{\perp \alpha
_2}}{k_1^{-}k_2^{+}}-g_{\perp }^{\alpha _1\alpha _2}
\end{equation}
and
$$
\gamma ^{\alpha _2\alpha _1\beta }(k_2,-k_1)\rightarrow \widetilde{\gamma }%
^{\alpha _2\alpha _1\beta }=(k_2-k_1)^\beta \left( g_{\perp }^{\alpha
_2\alpha _1}+2\frac{k_1^{\perp \alpha _1}k_2^{\perp \alpha _2}}{%
k_1^{-}k_2^{+}}\right)
$$
\begin{equation}
+2\left( k_1^{\perp }-\frac{k_1^{+}}{k_2^{+}}k_2^{\perp }\right) ^{\alpha
_2}\left( g_{\perp }^{\alpha _1\beta }-\frac{n^{-\beta }}{k_1^{-}}k_1^{\perp
\alpha _1}\right) -2\left( k_2^{\perp }-\frac{k_2^{-}}{k_1^{-}}k_1^{\perp
}\right) ^{\alpha _1}\left( g_{\perp }^{\alpha _2\beta }-\frac{n^{+\beta }}{%
k_2^{+}}k_2^{\perp \alpha _2}\right) .
\end{equation}

Thus, one can rewrite $A^{\alpha _2\alpha _1}$ as follows
\begin{equation}
A^{\alpha _1\alpha _2}\rightarrow a^{\alpha _1\alpha _2}=-\frac 12\frac{
\widetilde{\gamma }^{\alpha _2\alpha _1\beta }\Gamma ^{+-\beta }(q_2,q_1)}{%
(k_1+k_2)^2}+b^{\alpha _2\alpha _1},
\end{equation}
where
\begin{equation}
b^{\alpha _1\alpha _2}=-2\,\frac{k_1^{-}k_2^{+}g_{\perp }^{\alpha _2\alpha
_1}-2Q_1^{\perp \alpha _1}Q_1^{\perp \alpha _2}}t+2\frac{k_1^{\perp \alpha
_1}k_2^{\perp \alpha _2}}{k_1^{-}k_2^{+}}-g_{\perp }^{\alpha _1\alpha _2}
\end{equation}
and
\begin{equation}
Q_1=q_1-k_1=q_2+k_2\,,\,\,t=Q_1^2.
\end{equation}

In an explicit form the tensor $a^{\alpha _1\alpha _2}$ after the suitable
renormalization is given below (see $\left[ 14\right] $):
$$
c^{\alpha _1\alpha _2}\equiv \frac 14a^{\alpha _1\alpha _2}=\frac{Q_{1\perp
}^{\alpha _1}Q_{1\perp }^{\alpha _2}}t-\frac{Q_{1\perp }^{\alpha _1}}\kappa
(k_{1\perp }^{\alpha _2}-\frac{k_1^{+}}{k_2^{+}}k_{2\perp }^{\alpha _2})+
\frac{Q_{1\perp }^{\alpha _2}}\kappa (k_{2\perp }^{\alpha _1}-\frac{k_2^{-
}}{%
k_1^{-}}k_{1\perp }^{\alpha _2})
$$
$$
+\frac{k_{1\perp }^{\alpha _1}k_{1\perp }^{\alpha _2}}\kappa \frac{q_{2\perp
}^2}{k_1^{-}(k_1^{+}+k_2^{+})}+\frac{k_{2\perp }^{\alpha _1}k_{2\perp
}^{\alpha _2}}\kappa \frac{q_{1\perp }^2}{k_2^{+}(k_1^{-}+k_2^{-})}-\frac{%
k_{1\perp }^{\alpha _1}k_{2\perp }^{\alpha _2}}\kappa \left( 1+\frac
t{k_1^{-}k_2^{+}}\right) +\frac{k_{2\perp }^{\alpha _1}k_{1\perp }^{\alpha
_2}}\kappa
$$
\begin{equation}
-\frac{g_{\perp }^{\alpha _1\alpha _2}}2\left( 1+\frac t\kappa +\frac{%
k_2^{+}k_1^{-}}t+\frac{k_2^{+}k_1^{-}-k_2^{-}k_1^{+}}\kappa -\frac{q_{1\perp
}^2}\kappa \frac{k_2^{-}}{k_1^{-}+k_2^{-}}-\frac{q_{2\perp }^2}\kappa \frac{%
k_1^{+}}{k_1^{+}+k_2^{+}}\right) ,
\end{equation}
where $\kappa =(k_1+k_2)^2.$

The amplitude of producing a pair of massless quark and antiquark with their
momenta $k_1 $ and $k_2$ correspondingly also can be written as a sum of two
terms being the matrices in the spin and colour spaces:

\begin{equation}
\psi _{c_2c_1}=-g^2\left(
t^{c_1}t^{c_2}b(k_1,k_2)-t^{c_2}t^{c_1}b^T(k_2,k_1)\right) ,
\end{equation}
where $t^c$ are the colour group generators in the fundamental
representation and the expressions for $b(k_1,k_2)$ and $b^T(k_1,k_2)$ are
constructed according to the Feynman rules including the effective vertices
(see the diagrams of Fig.4 describing $b(k_1,k_2)$) [14]:
$$
b(k_1,k_2)=\gamma ^{-}\frac{\widehat{q_1^{\perp }}-\widehat{k_1^{\perp }}}{%
(q_1-k_1)^2}\gamma ^{+}-\frac{\gamma ^\beta \Gamma ^{+-\beta }(q_2,q_1)}{%
(k_1+k_2)^2}\,,
$$
\begin{equation}
b^T(k_2,k_1)=\gamma ^{+}\frac{\widehat{q_1^{\perp }}-\widehat{k_2^{\perp
}}}{%
(q_1-k_2)^2}\gamma ^{-}-\frac{\gamma ^\beta \Gamma ^{+-\beta }(q_2,q_1)}{%
(k_1+k_2)^2}\,.
\end{equation}

We have the following relations valid for $q_1^{\perp }\rightarrow 0$ :

\begin{equation}
\Gamma ^{+-\beta }(q_2,q_1)\rightarrow 2(k_1+k_2)^\beta +2(k_1+k_2)^2\frac{%
n^{+\beta }}{k_1^{+}+k_2^{+}}\,,\,\,(q_1-k_i)^2\rightarrow
-k_i^{-}(k_1^{+}+k_2^{+})\,.
\end{equation}
Therefore using the Dirac equations for the spinors:

\begin{equation}
\overline{u}(k_1)\widehat{k_1}=\widehat{k_2}v(k_2)=0
\end{equation}
one obtains that the quark production amplitude vanishes in the limit of
small $q_1^{\perp }$ or $q_2^{\perp }$:

\begin{equation}
\overline{u}(k_1)\psi _{c_2c_1}v(k_2)\,\rightarrow 0\,
\end{equation}
analogously to the gluon case (see below). It is convenient to present the
wave functions $u(k_1)$ and $v(k_2)$ of the produced quark and antiquark
correspondingly as linear combinations of four definite spinors $\left[
5\right] $

\begin{equation}
u(k_1)=\sum_ic_i(k_1)u_i\,,v(k_2)=\sum_id_i(k_2)u_i\,,
\end{equation}
where
\begin{equation}
u_{--}=\frac 12\left(
\begin{array}{c}
0 \\
1 \\
0 \\
-1
\end{array}
\right) ,\,u_{+-}=\frac 12\left(
\begin{array}{c}
0 \\
1 \\
0 \\
1
\end{array}
\right) ,\,u_{-+}=-\frac 12\left(
\begin{array}{c}
1 \\
0 \\
1 \\
0
\end{array}
\right) ,\,u_{++}=\frac 12\left(
\begin{array}{c}
1 \\
0 \\
-1 \\
0
\end{array}
\right) \,.
\end{equation}

These spinors have the properties:

$$
\gamma _{+}u_{-k}=0,\,\gamma _{-}u_{+k}=0,\,\gamma \,u_{k+}=0,\,\gamma
^{*}u_{k-}=0\,,
$$

$$
\gamma \,u_{--}=-\gamma _{+}u_{++}=2u_{-+}\,,\,\gamma ^{*}u_{++}=-\gamma
_{-}u_{--}=-2u_{+-}\,,
$$

\begin{equation}
\gamma \,u_{+-}=-\gamma _{-}u_{-+}=2u_{++}\,,\,\gamma ^{*}u_{-+}=-\gamma
_{+}u_{+-}=-2u_{--}\,,
\end{equation}
where
\begin{equation}
\gamma _{\pm }=\gamma _0\pm \gamma _3\,,\,\gamma =\gamma ^1+i\gamma
^2\,,\,\gamma *=\gamma ^1-i\gamma ^2\,,\gamma _0=\left(
\begin{array}{cc}
1 & 0 \\
0 & -1
\end{array}
\right) \,,\,\gamma ^i=\left(
\begin{array}{cc}
0 & \sigma _i \\
-\sigma _i & 0
\end{array}
\right)
\end{equation}
and $\sigma _i$ are the Pauli matrices. Note, that $\gamma^{\pm}=\gamma^0
\pm \gamma^3$ and $\gamma^0 =\gamma_0,\, \gamma^i=-\gamma_i$. The quark and
anti-quark eigenstates of the matrix $\gamma _5$:
\begin{equation}
\gamma _5=\left(
\begin{array}{cc}
0 & 1 \\
1 & 0
\end{array}
\right) =-\frac 14(\gamma _{+}\gamma _{-}-\gamma _{-}\gamma _{+})\,\frac
14(\gamma \gamma ^{*}-\gamma^*\gamma )
\end{equation}
are described correspondingly by the spinors $u^{(\pm )}(k_1)$ and $%
\,v^{(\pm )}(k_2)$ with positive and negative helicities $\lambda =\pm \frac
12$. They can be written as follows:
$$
u^{(+)}(k_1)=c_{-+}(k_1)\,u_{-+}+c_{+-}(k_1)\,u_{+-}\,,\,%
\,v^{(-)}(k_2)=d_{-+}(k_2)\,u_{-+}+d_{+-}(k_2)\,u_{+-}\,;
$$
\begin{equation}
u^{(-)}(k_1)=c_{++}(k_1)\,u_{++}+c_{--}(k_1)\,u_{--}\,,\,%
\,v^{(+)}(k_2)=d_{++}(k_1)\,u_{++}+d_{--}(k_2)\,u_{--}\,.
\end{equation}

Due to the Dirac equation for $u(k_1)$ and $v(k_2)$ the coefficients $c_i$
and $d_i$ are not independent:%
$$
c_{+-}(k_1)=-\frac{k_1^{-}}{k_1^{*}}c_{-+}=-\frac{k_1}{k_1^{+}}%
c_{-+}\,,\,d_{+-}(k_2)=-\frac{k_2^{-}}{k_2^{*}}d_{-+}=-\frac{k_2}{k_2^{+}}%
d_{-+}\,,
$$
\begin{equation}
c_{--}(k_1)=-\frac{k_1}{k_1^{-}}c_{++}=-\frac{k_1^{+}}{k_1^{*}}%
c_{++}\,,\,d_{--}(k_2)=-\frac{k_2}{k_2^{-}}d_{++}=-\frac{k_2^{+}}{k_2^{*}}%
d_{++}\,,
\end{equation}
where we use the complex components $k=k^1+ik^2,\,k^*=k^1-ik^2$ of two
dimensional vector $k^\alpha$ .

In accordance with the normalization condition for the wave functions

\begin{equation}
\overline{u}(k_1)\gamma ^\alpha u(k_1)=2k_1^\alpha \,,\,\overline{v}%
(k_2)\gamma ^\alpha v(k_2)=2k_2^\alpha
\end{equation}
we have the other constraints on these coefficients:

\begin{equation}
\mid c_{\pm i}(k_1)\mid ^2=2k_1^{\mp }\,,\,\mid d_{\pm i}(k_2)\mid
^2=2k_2^{\mp }\,.
\end{equation}

Using above relations one can calculate the matrix elements of the
production amplitude:

$$
\overline{u}^{(+)}(k_1)b(k_1,k_2)v^{(-)}(k_2)=\frac
12c_{+-}^{*}(k_1)d_{-+}(k_2)\,b^{(+-)}(k_1,k_2)\,,
$$
$$
\overline{u}^{(-)}(k_1)b^T(k_2,k_1)v^{(-)}(k_2)=\frac{1}{2}c_{-+}^*(k_1)
d_{+-}(k_2)\,b^{(-+)}(k_2,k_1)\,,
$$
$$
\overline{u}^{(-)}(k_1)b(k_1,k_2)v^{(+)}(k_2)=\frac
12c_{++}^{*}(k_1)d_{--}(k_2)\,b^{(-+)}(k_1,k_2)\,,
$$
\begin{equation}
\overline{u}^{(-)}(k_1)b^T(k_2,k_1)v^{(+)}(k_2)=\frac{1}{2}%
c_{--}^*(k_1)d_{++} (k_2)\,b^{(+-)}(k_2,k_1)\,,
\end{equation}
where

\begin{equation}
b^{(+-)}(k_1,k_2)=(b^{(-+)}(k_1,k_2))^{*}=-4\,\frac{q_1-k_1}{(q_1-k_1)^2}-
\frac{j^\beta \Gamma ^{+-\beta }(q_2,q_1)}{(k_1+k_2)^2}\,,
\end{equation}
and the quark current $j$ is
\begin{equation}
j=n+n^{*}\,\frac{k_2^{-}}{k_2^{*}}\,\frac{k_1}{k_1^{-}}-n^{-}\,\frac{k_1}{%
k_1^{-}}-n^{+}\,\frac{k_2^{-}}{k_2^{*}}\,.
\end{equation}
By definition we have
$$
n^\alpha k^\alpha =k,\,n^{*\alpha }k^\alpha =k^{*}\,,\,n^{\pm \alpha
}k^\alpha =k^{\pm }.
$$
This current is conserved:
\begin{equation}
(k_1^\beta +k_2^\beta )\,j^\beta =0\,.
\end{equation}

\section{Final state particles with definite helicities and the complex
transverse momenta}

Using the reality condition $s\alpha _i\beta _i=\overrightarrow{k_i}^2$ for
the produced gluons $i=1,2$ to exclude the Sudakov parameters $\alpha _{i
}$one can present the tensor $c^{\alpha _1\alpha _2}$ only in terms
of transverse momenta $\overrightarrow{k_i}\,,\,\overrightarrow{q_i}$ and
the relative parameter $x=\frac{\beta _1}{\beta _1+\beta _2}$:

$$
c^{\alpha _1\alpha _2}=\frac{\delta ^{\alpha _1\alpha _2}}{2Z}
\overrightarrow{q_1}^2x(1-x)-xk_1^{\alpha _1}\frac{\overrightarrow{q_2}%
^2k_1^{\alpha _2}+\overrightarrow{\Delta }^2Q_1^{\alpha _2}-(1-x)^{-1}(
\overrightarrow{Q_1}^2-\overrightarrow{k_1}^2)k_2^{\alpha _2}}{\kappa
\overrightarrow{\,k_1}^2}-
$$
$$
-\frac 1{\overrightarrow{k_1}^2}xk_1^{\alpha _1}Q_1^{\alpha _2}+\frac{\Delta
^{\alpha _1}q_1^{\alpha _2}+\delta ^{\alpha _1\alpha _2}\overrightarrow{q_1}%
( \overrightarrow{k_1}+x\overrightarrow{q_2})-(1-x)^{-1}q_1^{\alpha
_1}(k_1^{\alpha _2}-x\Delta ^{\alpha _2})}\kappa +
$$
\begin{equation}
+\frac{Q_1^{\alpha _1}Q_1^{\alpha _2}-\frac 12(1-x)(\overrightarrow{Q_1}^2-
\overrightarrow{k_1}^2)\delta ^{\alpha _1\alpha _2}}t+x\overrightarrow{q_1}%
^2\,\frac{k_2^{\alpha _1}k_2^{\alpha _2}+\delta ^{\alpha _1\alpha _2}(1-x)
\overrightarrow{k_1}\overrightarrow{\Delta }}{\kappa Z}
\end{equation}
where $\delta ^{\alpha _1\alpha _2}=-g_{\perp }^{\alpha _1\alpha _2}$ is the
Kroniker tensor ($\delta ^{11}=\delta ^{22}=1$) and

$$
x\equiv \frac{k_1^{+}}{k_1^{+}+k_2^{+}}\,,\,\overrightarrow{\Delta }\equiv
\overrightarrow{q_1}-\overrightarrow{q_2}=\overrightarrow{k_1}+
\overrightarrow{k_2}\,,\,\overrightarrow{Q_1}=\overrightarrow{q_1}-
\overrightarrow{k_1}\,,\,\kappa =\frac{(\overrightarrow{k_1}-x
\overrightarrow{\Delta })^2}{x(1-x)}\,,
$$
\begin{equation}
Z=-x(1-x)(\overrightarrow{\Delta }^2+\kappa )\,,\,\,t=-\frac{(
\overrightarrow{k_1}-x\overrightarrow{q_1})^2+x(1-x)\overrightarrow{q_1}^2}%
x\,.
\end{equation}
The imaginary part of the elastic scattering amplitude calculated with the
use of the $s$-channel unitarity condition in terms of the squared
production amplitude $A_{2\rightarrow 4}$ contains the infrared divergences
at small $k_{i\perp }^2$ and $\kappa $. To avoid such divergencies the
dimensional regularization

\begin{equation}
\frac{d^4k}{(2\pi )^4}\,\longrightarrow \frac{d^Dk}{(2\pi )^D}
\end{equation}
is used in the gauge theories. In the previous papers $\left[ 11-13\right] $
we calculated in this regularization the next to leading corrections to the
production amplitude in the multi-Regge kinematics for the final particles.
In the $D$-dimensional space the gluon has $D-2$ degrees of freedom. In our
case the transverse momenta $\overrightarrow{q_1},\overrightarrow{q_2}$ of
the reggeized gluons form a plane in the $D-2$ -dimensional space for the
polarization vectors $\overrightarrow{e_{\perp }}$. Two states described by
the vectors $\overrightarrow{e_{\perp }}$ belonging to this plane can be
considered as the physical states. Other $D-4$ states are the auxilliary
states necessary for the regularization. For $D\rightarrow 4$ their
contribution in the region of fixed $k_{i\perp }$ and $\kappa $ becomes
negligible.

We begin with the physical contributions from the region of non-vanishing $%
k_{i\perp }$ and $\kappa $ where the transverse subspace can be considered
as two-dimensional one. The singular region of small $k_{i\perp }$ and $%
\kappa $ will be discussed later. Let us introduce the complex components of
the tensor $c^{\alpha _1\alpha _2}$ directly related with the amplitudes for
producing the gluons with equal and opposite helicities correspondingly
(independently this result was obtained also in ref. [16]):

$$
c^{+-}(k_1,k_2)\equiv c^{11}+ic^{21}-ic^{12}+c^{22}=-\frac{%
q_2\,(k_1^{*}-x\Delta ^{*})\,q_1^{*}}{\kappa \,k_1^{*}(1-x)}\,,
$$
$$
c^{++}(k_1,k_2)=c^{11}+ic^{21}+ic^{12}-c^{22}=\frac{Q_1Q_1}t-\frac{xQ_1}{%
k_1^{*}}+\frac{x\overrightarrow{q_1}^2k_2}\kappa \left( \frac{k_2}Z+\frac
1{k_1^{*}}\right) -
$$
\begin{equation}
-\frac{xq_1(k_1-x\Delta )}{\kappa \,(1-x)}-\frac 1{\kappa \,k_1^{*}}\left(
(k_1-x\Delta )\frac x{1-x}(\overrightarrow{Q_1}^2-\overrightarrow{k_1}%
^2)-(k_1^{*}-x\Delta ^{*})q_1k_2\right) ,
\end{equation}
where $B=B^1+i\,B^2,\,B^{*}=B^1-iB^2$ are complex coordinates of a
two-dimensional vector $\overrightarrow{B}$. The above expression for $%
c^{\alpha _1\alpha _2}$ can be presented in the following form:

$$
c^{+-}(k_1,k_2)=\overline{c^{-+}}(k_1,k_2)=-x\,\frac{q_2\,q_1^{*}}{%
(k_1-x\,\Delta )\,k_1^{*}}\,,
$$

$$
c^{++}(k_1,k_2)=\overline{c^{--}}(k_1,k_2)=
$$
$$
=-\frac{x\,(Q_1)^2}{\left( (\overrightarrow{k_1}-x\overrightarrow{q_1}%
)^2+x(1-x)\overrightarrow{q_1}^2\right) }+\frac{x\,\overrightarrow{q_1}%
^2(k_2)^2}{\overrightarrow{\Delta }^2\left( (\overrightarrow{k_1}-x
\overrightarrow{\Delta })^2+x(1-x)\overrightarrow{\Delta }^2\right) }-
$$
\begin{equation}
-\frac{x(1-x)q_1k_2q_2^{*}}{\Delta ^{*}\,(k_1-x\Delta )k_1^{*}}-\frac{%
xq_1^{*}k_1q_2}{\overrightarrow{\Delta }^2(k_1^{*}-x\Delta ^{*})}+\frac{%
xq_2^{*}Q_1}{\Delta ^{*}\,k_1^{*}}\,.
\end{equation}
One can verify, that these expressions have the following symmetry
properties:
\begin{equation}
c^{+-}(k_1,k_2) \leftrightarrow c^{-+}(k_1,k_2),\,c^{++}(k_1,k_2)
\leftrightarrow c^{++}(k_1,k_2)
\end{equation}
under the simultaneous substitutions
$$
k_1\leftrightarrow k_2\,,\,q_1\leftrightarrow -q_2\,,\,x\leftrightarrow
\frac{x\overrightarrow{k_2}^2}{(\overrightarrow{k_1}-x\overrightarrow{\Delta
})^2+x(1-x)\overrightarrow{\Delta }^2}\,,
$$
\begin{equation}
k_1-x\Delta \leftrightarrow \frac{k_1k_2(k_1^{*}-x\Delta ^{*})}{(
\overrightarrow{k_1}-x\overrightarrow{\Delta })^2+x(1-x)\overrightarrow{%
\Delta }^2},
\end{equation}
corresponding to the left-right symmetry:
\begin{equation}
k_1\leftrightarrow k_2\,,\,q_1\leftrightarrow -q_2\,,\,n^{+}\leftrightarrow
n^{-}.
\end{equation}

In the Regge regime of small $1-x$ and fixed $k_i,q_i$ amplitudes $c^{\alpha
_1\alpha _2}$ are simplified as follows
\begin{equation}
c^{+-}(k_1,k_2)\longrightarrow \frac{q_1^{*}\,q_2}{k_1^{*}\,k_2}%
\,,\,\,c^{++}(k_1,k_2)\longrightarrow \frac{q_1^{*}}{k_1^{*}}\,\frac{q_1-k_1
}{q_1^{*}-k_1^{*}}\,\frac{q_2^{*}}{k_2^{*}}
\end{equation}
and are proportional to the product of the effective vertices $\Gamma
^{+-\beta }$ in the light-cone gauge $\left[ 4\right] $. For $x\rightarrow 0$
the amplitudes $c^{+-}(k_1,k_2)$ and $c^{++}(k_1,k_2)$ vanish, but for
simultaneously small $k_1$ or $k_1-x\Delta $ one obtains from them a nonzero
integral contribution because in this region there are poles:

\begin{equation}
c^{++}(k_1,k_2)\rightarrow \frac{q_1q_2^{*}\Delta }{q_1^{*}q_2\Delta ^{*}}%
\,c^{+-}(k_1,k_2)\rightarrow -\frac{xq_1q_2^{*}\Delta }{\Delta
^{*}(k_1-x\Delta )k_1^{*}}\,.
\end{equation}
For large $k_1$ and fixed $q_i\,,\,x$ we obtain
\begin{equation}
c^{+-}(k_1,k_2)\longrightarrow -x\,\frac{q_1^{*}q_2}{\overrightarrow{k_1}^2}%
\,,\,\,c^{++}(k_1,k_2)\longrightarrow -x(1-x)^2\,\frac{q_1q_2}{
\overrightarrow{k_1}^2}-x^3\frac{q_1^{*}q_2^{*}k_1}{(k_1^{*})^3}
\end{equation}
and therefore the integrals for the cross section of the gluon production do
not contain any ultraviolet divergency.

As it is seen from above formulas, $c^{\alpha _1\alpha _2}$ contains
infrared poles at small $\overrightarrow{k_1}$:
\begin{equation}
c^{+-}(k_1,k_2)\rightarrow \,\frac{q_1^{*}q_2}{\Delta \,k_1^{*}}%
\,,\,c^{++}(k_1,k_2)\rightarrow \,\frac{q_1q_2{*}}{\Delta ^{*}\,k_1^{*}}\,\,
\end{equation}
and at small $\overrightarrow{k_1}-x\,\overrightarrow{\Delta }$:
\begin{equation}
c^{+-}\rightarrow -\,\frac{q_1^{*}q_2}{\Delta ^{*}(k_1-x\Delta )}%
\,,\,\,c^{++}\rightarrow -(1-x)^2\,\frac{q_1\Delta q_2^{*}}{(\Delta
^{*})^2(k_1-x\Delta )}-x^2\frac{q_1^{*}q_2}{\Delta ^{*}(k_1^{*}-x\Delta
^{*})%
}\,.
\end{equation}

It is obvious, that the amplitude $c^{+-}(k_1,k_2)$ vanishes for $%
q_1\rightarrow 0$ or $q_2\rightarrow 0$. The amplitude $c^{++}(k_1,k_2)$
behaves as follows
\begin{equation}
c^{++}(k_1,k_2)\rightarrow \frac{xk_1}{k_1^*} \left(\frac{x^2q_1^*\Delta^*}{%
k_1^*(k_1^*-x\Delta^*)}+ \frac{(1-x)^2q_1\Delta}{k_1(k_1-x\Delta )}\right)
\end{equation}
for $q_1\rightarrow 0$ and as follows

\begin{equation}
c^{++}(k_1,k_2)\rightarrow -\frac{x}{(\overrightarrow{k_1}^2-2x
\overrightarrow{k_1} \overrightarrow{\Delta}+x\overrightarrow{\Delta}^2)^2}
\left(\frac{x^2q_2^*\Delta^*k_2^4}{k_1^*(k_1-x\Delta)}+ \frac{%
(1-x)^2q_2\Delta\overrightarrow{k_1}^2k_1^*}{k_1^*-x\Delta^*}\right)
\end{equation}
for $q_2\rightarrow 0$. The last expression can be obtained from previous
one with the use of the above left-right symmetry of $c^{++}(k_1,k_2)$.

Let us return now to the quark-anti-quark production and renormalize the
amplitude as follows
\begin{equation}
b^{(+-)}(k_1,k_2)=\frac 4{k_1^{*}}\,c(k_1,k_2)\,.
\end{equation}
Then the factor $c(k_1,k_2)$ is given below:
$$
c(k_1,k_2)=\frac{xQ_1k_1^{*}}{(\overrightarrow{k_1}-x\overrightarrow{q_1}%
)^2+x(1-x)\overrightarrow{q_1}^2}-\frac{x\overrightarrow{q_1}^2k_2k_1^{*}}{
\overrightarrow{\Delta }^2\left( (\overrightarrow{k_1}-x\overrightarrow{%
\Delta })^2+x(1-x)\overrightarrow{\Delta }^2\right) }
$$
\begin{equation}
+\frac{x(1-x)q_1q_2^{*}}{\Delta ^{*}(k_1-x\Delta )}-\frac{xq_1^{*}q_2k_1^{*}
}{\overrightarrow{\Delta }^2(k_1^{*}-x\Delta ^{*})}-\frac{xq_2^{*}}{\Delta
^{*}}.
\end{equation}
It can be written as follows
$$
\frac 1x\,c(k_1,k_2)=\frac{(1-x)q_1k_1^{*}-xq_1^{*}k_1+x\overrightarrow{q_1}%
^2}{(\overrightarrow{k_1}-x\overrightarrow{q_1})^2+x(1-
x)\overrightarrow{q_1}%
^2}
$$
\begin{equation}
-\frac{\overrightarrow{q_1}^2}{\overrightarrow{\Delta }^2}\,\frac{%
(1-x)\Delta k_1^{*}-x\Delta ^{*}k_1+x\overrightarrow{\Delta }^2}{(
\overrightarrow{k_1}-x\overrightarrow{\Delta })^2+x(1-x)\overrightarrow{%
\Delta }^2}+\frac{(1-x)q_1q_2^{*}}{\Delta ^{*}(k_1-x\Delta )}-\frac{%
xq_1^{*}q_2}{\Delta (k_1^{*}-x\Delta ^{*})}\,.
\end{equation}
For large values of $k_1$ this expression decreases rapidly:
\begin{equation}
\frac 1xc(k_1,k_2)\rightarrow x(1-x)\frac{q_1q_2}{k_1^2}-x^2\frac{%
q_1^{*}q_2^{*}}{(k_1^{*})^2}\,.
\end{equation}
The amplitude $c(k_1,k_2)$ vanishes

\begin{equation}
c(k_1,k_2)\rightarrow \frac{x^3q_1^{*}\Delta ^{*}}{k_1^{*}(k_1^{*}-x\Delta
^{*})}-\frac{x^2(1-x)q_1\Delta }{k_1(k_1-x\Delta )}
\end{equation}
for $\overrightarrow{q_1}\rightarrow 0$ and

\begin{equation}
c(k_1,k_2)\rightarrow \frac{x^3k_2^3q_2^{*}\Delta ^{*}(k_1^{*}-x\Delta
^{*})-x^2(1-x)(k_1^{*})^2k_2^{*}q_2\Delta (k_1-x\Delta )}{\left( (
\overrightarrow{k_1}-x\overrightarrow{\Delta })^2+x(1-x)\overrightarrow{%
\Delta }^2\right)^2 (\overrightarrow{k_1}-x\overrightarrow{\Delta })^2}
\end{equation}
for $\overrightarrow{q_2}\rightarrow 0$ correspondingly. The last expression
can be obtained from previous one if we take into account that under above
left-right symmetry transformations $c(k_1,k_2)$ is changed as follows

\begin{equation}
c(k_1,k_2)\leftrightarrow -\frac{k_2^{*}}{k_1^{*}}c(k_1,k_2)\,.
\end{equation}

\section{Squared production amplitudes}

The total cross-section of the gluon production is proportional to the
integral from the squared
amplitude $\psi _{d_1d_2c_2c_1}^{\alpha _1\alpha _2}$ summed over all
indices. It is expressed in terms of the quantity

\begin{equation}
R\equiv (c^{\alpha _1\alpha _2}(k_1,k_2))^2+(c^{\alpha _1\alpha
_2}(k_2,k_1))^2+c^{\alpha _1\alpha _2}(k_1,k_2)c^{\alpha _2^{\prime }\alpha
_1^{\prime }}(k_2,k_1)\,\Omega ^{\alpha _1\alpha _1^{\prime }}\, \Omega
^{\alpha_2\alpha _2^{\prime }},
\end{equation}
where the tensor $\Omega ^{\alpha _i\alpha _{i^{\prime }}}$ interchanges the
left and right gauges (see $\left[ 4\right] $):

$$
\Omega ^{\alpha _i\alpha _{i^{\prime }}}=g_{\perp }^{\alpha _i\alpha
_{i^{\prime }}}-2\frac{k_{i\perp }^{\alpha _i}k_{i\perp }^{\alpha
_{i^{\prime }}}}{k_{i\perp }^2}\,.
$$
Above we took into account that the colour factor for the interference term
is two times smaller than one for the direct contributions. The generalized
BFKL equation for the virtual gluon cross-section can be written in the
integral form as follows
\begin{equation}
\sigma (\overrightarrow{q_1},\,q_1^{+})=\sigma _0(\overrightarrow{q_1}%
,q_1^{+})+\int \frac{d\,q_2^{+}}{q_2^{+}}\int d^2\overrightarrow{q_2}%
\,K_\delta (\overrightarrow{q_1},\overrightarrow{q_2})\sigma (
\overrightarrow{q_2},\,q_2^{+})\,
\end{equation}
where the integration region over the longitudinal momentum $q_2^{+}$ is
restricted from above by the value proportional to $q_1^{+}$:
\begin{equation}
q_2^{+}\,<\,\delta \,q_1^{+}.
\end{equation}
The intermediate infinitesimal parameter $\delta >0$ is introduced to
arrange the particles in the groups with strongly different rapidities (see
$%
\left[ 14\right] $). The integral kernel $K_\delta (\overrightarrow{q_1},
\overrightarrow{q_2})$ takes into account the interaction among the
particles incide each group where $\delta $ plays role of the ultraviolet
cut-off in their relative rapidities. The kernel $K_\delta $ can be
calculated in the perturbation theory:
\begin{equation}
K_\delta (\overrightarrow{q_1},\overrightarrow{q_2})=\sum_{r=1}^\infty
\left( \frac{g^2}{2(2\pi )^{D-1}}\right) ^rK_\delta ^{(r)}(\overrightarrow{%
q_1},\overrightarrow{q_2}).
\end{equation}
The real contribution to the kernel in the leading logarithmic approximation
is proportional to the square of the effective vertex:
\begin{equation}
K_{real}^{(1)}(\overrightarrow{q_1},\overrightarrow{q_2})=-\frac{N_c}{4\,
\overrightarrow{q_1}^2\,\overrightarrow{q_2}^2}\left( \Gamma ^{+-\beta
}(q_2,q_1)\right) ^2=N_c\,\frac 4{(\overrightarrow{q_1}-\overrightarrow{q_2}%
)^2}.
\end{equation}
The corresponding virtual contribution is proportional to the gluon Regge
trajectory $\omega (-\overrightarrow{q}_1^2)$.

The next to leading term in $K_\delta $ related with the two gluon
production is given below
\begin{equation}
K_{gluons}^{(2)}=\int d\kappa \,d^Dk_1\,\delta (k_1^2)\,\delta (k_2^2)\,
\frac{16N_c^2R}{\overrightarrow{q_1}^2\overrightarrow{q}_2^2}=\frac{%
16\,N_c^2 }{2\overrightarrow{q_1}^2\overrightarrow{q_2}^2}\int_\delta
^{1-\delta } \frac{dx}{x(1-x)}\int d^{D-2}\overrightarrow{k_1}\,R\,.
\end{equation}

The limits in the integral over $x$ correspond to a restriction from above
for the invariant mass $\sqrt{\kappa }$ of the produced gluons. In the
solution of the generalized BFKL equation the dependence of $\delta $ should
disappear.

For the physical value $D=4$ of the space-time dimension one can express $R$
in terms of the contributions from the states with the definite helicities:

\begin{equation}
R=R(+-)+R(++)
\end{equation}
where
$$
R(+-)=\frac 12\left( \mid c^{+-}(k_1,k_2)\mid ^2+\mid c^{+-}(k_2,k_1)\mid
^2+Re\,c^{+-}(k_1,k_2)\,c^{-
+}(k_2,k_1)\frac{k_1^{*}}{k_1}\frac{k_2}{k_2^{*}}%
\right) \,,
$$
\begin{equation}
R(++)=\frac 12\left( \mid c^{++}(k_1,k_2)\mid ^2+\mid c^{++}(k_2,k_1)\mid
^2+Re\,c^{++}(k_1,k_2)\,c^{++}(k_2,k_1)\frac{k_1^{*}}{k_1}\frac{k_2^{*}}{k_2}
\right) .
\end{equation}
Here we have used the following relation between the polarization vectors in
the right and left gauges $\left[ 4\right] $:
\begin{equation}
e_{\perp }^r(k)=-(e_{\perp }^l(k))^{*}\frac k{k^{*}}
\end{equation}
to express all helicity amplitudes in terms of two complex functions $%
c^{+-}(k_1,k_2)$ and $c^{++}(k_1,k_2)$ given above. One should also take
into account the following relations:

$$
c^{+-}(k_1,k_2)\,\frac{k_1^{*}}{k_1}=-\frac{q_2q_1^{*}}{\Delta }\,\left(
\frac 1{k_1-x\Delta }-\frac 1{k_1}\right) \,,
$$

$$
c^{++}(k_1,k_2)\,\frac{k_1^{*}}{k_1}=-\frac{(1-x)^2q_1q_2^{*}}{\Delta
^{*}(k_1-x\Delta )}-\frac{x^2q_1^{*}q_2}{\Delta (k_1^{*}-x\Delta ^{*})}+
\frac{(1-x)^2q_1q_2^{*}}{\Delta ^{*}k_1}
$$
$$
+\frac{x^2k_1q_1^{*}\left( (3-x)q_1-k_1\right) -xq_1^2\left(
x(2-x)q_1^{*}+(1-x)^2k_1^{*}\right) }{k_1\left( (\overrightarrow{k_1}-x
\overrightarrow{q_1})^2+x(1-x)\overrightarrow{q_1}^2\right) }
$$
\begin{equation}
+\frac{x^2\,\overrightarrow{q_1}^2\Delta ^{*}k_1\left( k_1-(3-x)\Delta
\right) +x\overrightarrow{q_1}^2(\Delta )^2\left( x(2-x)\Delta
^{*}+(1-x)^2k_1^{*}\right) }{\overrightarrow{\Delta }^2k_1\left( (
\overrightarrow{k_1}-x\overrightarrow{\Delta })^2+x(1-x)\overrightarrow{%
\Delta }^2\right) }.
\end{equation}
The various bilinear combinations of the functions $c^{+-}$ and $c^{++}$
needed to calculate $R(+-)$ and $R(++)$ are given in Appendix. The
ultraviolet divergencies at large $k_1$ which appear in some contributions
are cancelled in the sum.

At small $k_1$ and fixed $x$ we obtain:

\begin{equation}
\mid c^{++}(k_1,k_2)\mid ^2\rightarrow \mid c^{+-}(k_1,k_2)\mid
^2\rightarrow \frac{\overrightarrow{q_1}^2\overrightarrow{q_2}^2}{
\overrightarrow{k_1}^2\overrightarrow{\Delta }^2}\,.\,\,
\end{equation}
In this region the interference terms are convergent. The quantity $R$ at
small $\overrightarrow{k_1}$ and fixed $x$ can be calculated in the $D$%
-dimensional space-time:

\begin{equation}
R\rightarrow \frac{\overrightarrow{q_1}^2\overrightarrow{q_2}^2}{
\overrightarrow{\Delta }^2\overrightarrow{k_1}^2}\,,
\end{equation}
which gives a possibility to regularize the infrared divergency at $k_1
\rightarrow 0$.

At small $k_1-x\Delta $ and fixed $x$ we obtain for the bilinear
combinations of $c^{+-}$:
\begin{equation}
\mid c^{+-}(k_1,k_2)\mid ^2\rightarrow -c^{+-}(k_1,k_2)\,c^{-+}(k_2,k_1)\,
\frac{k_1^{*}}{k_1}\frac{k_2}{k_2^{*}}\rightarrow \frac{\overrightarrow{q_1}%
^2\overrightarrow{q_2}^2}{\overrightarrow{\Delta }^2(\overrightarrow{k_1}-x
\overrightarrow{\Delta })^2}\,.
\end{equation}
The bilinear combinations of $c^{++}$ in the same region are simplified
drastically:

\begin{eqnarray*}
\mid c^{++}(k_1,k_2)\mid ^2\rightarrow -c^{++}(k_1,k_2)c^{++}(k_2,k_1)\frac{%
k_1^{*}}{k_1}\frac{k_2^{*}}{k_2}\rightarrow \mid \frac{(1-x)^2q_1q_2^{*}}{%
\Delta ^{*}(k_1-x\Delta )}+\frac{x^2q_1^{*}q_2}{\Delta (k_1^{*}-x\Delta
^{*})%
}\mid ^2
\end{eqnarray*}
\begin{equation}
=\frac{\overrightarrow{q_1}^2\overrightarrow{q_2}^2(1-2x)^2}
{\overrightarrow{\Delta }^2(\overrightarrow{k_1}-
x \overrightarrow{\Delta })^2}\,
+ \frac{4x^2(1-x)^2(\overrightarrow{q_1}^2\overrightarrow{\Delta }-
\overrightarrow{\Delta }^2
\overrightarrow{q_1}\,,\,\overrightarrow{k_1}-x\overrightarrow{\Delta })^2 }
{\overrightarrow{\Delta }^4(\overrightarrow{k_1}-x \overrightarrow{\Delta })
^4}.
\end{equation}
The quantity $R$ at small $\overrightarrow{k_1}-x\overrightarrow{\Delta }$
and fixed $x$ can be calculated for an arbitrary dimension $D$ of the
space-time if we take into account that the tensor $c^{\alpha _1\alpha
_2}(k_1,k_2)$ is simplified in this limit:%
$$
c^{\alpha _1\alpha _2}(k_1,k_2)\rightarrow \widetilde{c}^{\alpha _1\alpha
_2}(k_1,k_2)=-2\frac{(1-x)\Delta ^{\alpha _1}\Delta ^{\alpha _2}}{
\overrightarrow{\Delta }^4}\,\frac{\left( (1-x)\overrightarrow{q_1}^2
\overrightarrow{\Delta }+x\overrightarrow{\Delta }^2\overrightarrow{q_1}%
\,,\, \overrightarrow{k_1}-x\overrightarrow{\Delta }\right) }{(
\overrightarrow{k_1}-x\overrightarrow{\Delta })^2}
$$
$$
+2\frac{(1-x)(\overrightarrow{\Delta },\,\overrightarrow{k_1}-x
\overrightarrow{\Delta })}{\overrightarrow{\Delta }^2(\overrightarrow{k_1}-x
\overrightarrow{\Delta })^2}\,\Delta ^{\alpha _1}q_1^{\alpha _2}-\frac{%
x(1-x)\,\delta ^{\alpha _1\alpha _2}}{\overrightarrow{\Delta }^2}\,\frac{(
\overrightarrow{q_1}^2\overrightarrow{\Delta }-\overrightarrow{\Delta }^2
\overrightarrow{q_1}\,,\,\overrightarrow{k_1}-x\overrightarrow{\Delta })}{(
\overrightarrow{k_1}-x\overrightarrow{\Delta })^2}
$$
\begin{equation}
+\frac{(1-x)(k_1^{\alpha _1}-x\Delta ^{\alpha _1})(\overrightarrow{q_1}%
^2\Delta ^{\alpha _2}-\overrightarrow{\Delta }^2q_1^{\alpha _2})}{
\overrightarrow{\Delta }^2(\overrightarrow{k_1}-x\overrightarrow{\Delta
})^2}%
-\frac{x(k_1^{\alpha _2}-x\Delta ^{\alpha _2})(\overrightarrow{q_2}^2\Delta
^{\alpha _1}+\overrightarrow{\Delta }^2q_2^{\alpha _1})}{\overrightarrow{%
\Delta }^2(\overrightarrow{k_1}-x\overrightarrow{\Delta })^2}\,.
\end{equation}
This expression has the important symmetry property:

\begin{equation}
\widetilde{c}^{\alpha _1\alpha _2}(k_1,k_2)=-\left( \delta ^{\alpha _1\alpha
_1^{\prime }}-2\frac{\Delta ^{\alpha _1}\Delta ^{\alpha _1^{\prime }}}{
\overrightarrow{\Delta }^2}\right) \left( \delta ^{\alpha _2\alpha
_2^{\prime }}-2\frac{\Delta ^{\alpha _2}\Delta ^{\alpha _2^{\prime }}}{
\overrightarrow{\Delta }^2}\right) \widetilde{c}^{\alpha _2^{\prime }\alpha
_1^{\prime }}(k_2,k_1)\,,
\end{equation}
which in particular means, that at $k\rightarrow x\Delta $ the interference
is destructive:
\begin{equation}
\widetilde{c}^{\alpha _1\alpha _2}(k_1,k_2)\,\widetilde{c}^{\alpha
_2^{\prime }\alpha _1^{\prime }}(k_2,k_1)\,\Omega ^{\alpha _1\alpha
_1^{\prime }}(\Delta )\,\Omega ^{\alpha _2\alpha _2^{\prime }}(\Delta
)=-\mid \widetilde{c}^{\alpha _1\alpha _2}(k_1,k_2)\mid ^2
\end{equation}
and
\begin{equation}
\mid \widetilde{c}^{\alpha _1\alpha _2}(k_2,k_1)\mid ^2=\mid \widetilde{c}%
^{\alpha _1\alpha _2}(k_1,k_2)\mid ^2\,.
\end{equation}
Let us write $\widetilde{c}$ as a sum of three terms:
\begin{equation}
\widetilde{c}^{\alpha _1\alpha _2}(k_1,k_2)=\frac{\delta ^{\alpha _1\alpha
_2}}{D-2}\,\widetilde{c}^{\alpha \alpha }(k_1,k_2)+\widetilde{c}^{\left[
\alpha _1\alpha _2\right] }(k_1,k_2)+\widetilde{c}^{(\alpha _1\alpha
_2)}(k_1,k_2)\,,
\end{equation}
where the trace of the matrix $\widetilde{c}^{\alpha _1\alpha _2}(k_1,k_2)$
is
\begin{equation}
\widetilde{c}^{\alpha \alpha }(k_1,k_2)=\left( (4-D)\,\frac{x(1-x)}{
\overrightarrow{\Delta }^2}\,\,\frac{\overrightarrow{q_1}^2\overrightarrow{%
\Delta }-\overrightarrow{\Delta
}^2\overrightarrow{q_1}}{(\overrightarrow{k_1%
}-x\overrightarrow{\Delta })^2}-\frac{\overrightarrow{q_2}^2\overrightarrow{%
\Delta }+\overrightarrow{\Delta }^2\overrightarrow{q_2}}{\overrightarrow{%
\Delta }^2(\overrightarrow{k_1}-x\overrightarrow{\Delta })^2}\,,\,
\overrightarrow{k_1}-x\overrightarrow{\Delta }\right) \,
\end{equation}
and
\begin{equation}
\widetilde{c}^{\left[ \alpha _1\alpha _2\right] }=\frac 12\left(
\widetilde{c%
}^{\alpha _1\alpha _2}-\widetilde{c}^{\alpha _2\alpha _1}\right) \,,\,
\widetilde{c}^{(\alpha _1\alpha _2)}=\frac 12\left( \widetilde{c}^{\alpha
_1\alpha _2}+\widetilde{c}^{\alpha _2\alpha _1}\right) -\frac{\delta
^{\alpha _1\alpha _2}}{D-2}\,\widetilde{c}^{\alpha \alpha }\,.
\end{equation}
These terms do not interfere each with others:
\begin{equation}
R=R^{(0)}+R^{(1)}+R^{(2)}\,.
\end{equation}
and for $k_1\rightarrow x\Delta $ their contributions are
\begin{equation}
R^{(0)}=\frac 1{D-2}\left( (4-D)\frac{x(1-x)}{\overrightarrow{\Delta }^2}(
\overrightarrow{q_1}^2\overrightarrow{\Delta }-\overrightarrow{\Delta }^2
\overrightarrow{q_1})-\frac{\overrightarrow{q_2}^2}{\overrightarrow{\Delta }%
^2}\overrightarrow{\Delta }-\overrightarrow{q_2}\,,\,\frac{\overrightarrow{%
k_1}-x\overrightarrow{\Delta }}{(\overrightarrow{k_1}-x\overrightarrow{%
\Delta })^2}\right) ^2
\end{equation}
for the singlet term,
$$
R^{(1)}=\frac 12\,\frac{\overrightarrow{q_1}^2\overrightarrow{q_2}^2}{
\overrightarrow{\Delta }^2(\overrightarrow{k_1}-x\overrightarrow{\Delta
})^2}%
-\frac 12\left( \frac{(\overrightarrow{q_2}^2\overrightarrow{\Delta }+
\overrightarrow{\Delta }^2\overrightarrow{q_2}\,,\,\overrightarrow{k_1}-x
\overrightarrow{\Delta })}{\overrightarrow{\Delta }^2(\overrightarrow{k_1}-x
\overrightarrow{\Delta })^2}\right) ^2
$$
\begin{equation}
+2\frac{x(1-x)}{\overrightarrow{\Delta }^2}\left( \left( \frac{(q_1^\beta
\overrightarrow{\Delta }-\Delta ^\beta \overrightarrow{q_1}\,,\,
\overrightarrow{k_1}-x\overrightarrow{\Delta })}{(\overrightarrow{k_1}-x
\overrightarrow{\Delta })^2}\right) ^2-\frac{\overrightarrow{q_1}^2
\overrightarrow{\Delta }^2-(\overrightarrow{\Delta },\overrightarrow{q_1})^2
}{(\overrightarrow{k_1}-x\overrightarrow{\Delta })^2}\right) \,
\end{equation}
for the antisymmetric tensor and
$$
R^{(2)}=-\frac 1{D-2}\,\left( \frac{\overrightarrow{q_2}^2\overrightarrow{%
\Delta }+\overrightarrow{\Delta }^2\overrightarrow{q_2}}{\overrightarrow{%
\Delta }^2}-2\frac{x(1-x)}{\overrightarrow{\Delta }^2}\left(
\overrightarrow{%
q_1}^2\overrightarrow{\Delta }-\overrightarrow{\Delta
}^2\overrightarrow{q_1}%
\right) \,,\,\frac{\overrightarrow{k_1}-x\overrightarrow{\Delta }}{(
\overrightarrow{k_1}-x\overrightarrow{\Delta })^2}\right) ^2
$$
$$
+4x^2(1-x)^2\left( \frac{\overrightarrow{q_1}^2\overrightarrow{\Delta }}{
\overrightarrow{\Delta }^2}-\overrightarrow{q_1}\,,\,\frac{\overrightarrow{%
k_1}-x\overrightarrow{\Delta }}{(\overrightarrow{k_1}-x\overrightarrow{%
\Delta })^2}\right) ^2+\frac 12\left( \frac{\overrightarrow{q_2}^2
\overrightarrow{\Delta }+\overrightarrow{\Delta }^2\overrightarrow{q_2}}{
\overrightarrow{\Delta }^2}\,,\,\frac{\overrightarrow{k_1}-x\overrightarrow{%
\Delta }}{(\overrightarrow{k_1}-x\overrightarrow{\Delta })^2}\right) ^2
$$
\begin{equation}
+\frac
12\frac{\overrightarrow{q_1}^2\overrightarrow{q_2}^2}{\overrightarrow{%
\Delta }^2(\overrightarrow{k_1}-x\overrightarrow{\Delta })^2}-2x(1-x)\left(
\frac{\overrightarrow{q_1}^2\overrightarrow{q_2}^2(\overrightarrow{\Delta },
\overrightarrow{k_1}-x\overrightarrow{\Delta })^2}{\overrightarrow{\Delta }%
^4(\overrightarrow{k_1}-x\overrightarrow{\Delta
})^4}+\frac{(\overrightarrow{%
q_1},\overrightarrow{q_2})^2}{\overrightarrow{\Delta
}^2(\overrightarrow{k_1}%
-x\overrightarrow{\Delta })^2}\right) .
\end{equation}
for the symmetric traceless tensor correspondingly. Note, that the total sum
$R$ is especially simple:

\begin{equation}
R=(D-2)\left( x(1-x)\frac{\overrightarrow{q_1}^2\overrightarrow{\Delta }-
\overrightarrow{\Delta }^2\overrightarrow{q_1}}{\overrightarrow{\Delta }^2}%
\,,\,\frac{\overrightarrow{k_1}-x\overrightarrow{\Delta }}{(\overrightarrow{%
k_1}-x\overrightarrow{\Delta })^2}\right) ^2+\frac{\left( 1-2x(1-x)\right)
\overrightarrow{q_1}^2\overrightarrow{q_2}^2}{\overrightarrow{\Delta }^2(
\overrightarrow{k_1}-x\overrightarrow{\Delta })^2}.
\end{equation}
We shall use this fact in the next section.

In the physical case $D=4$ the quantities $R^{(i)}$ can be expressed in
terms of the contributions from the definite helicity states:

$$
R(+-)\rightarrow R^{(0)}+R^{(1)}\,,
$$
$$
R^{(0)}=\frac 18\left| \frac{q_1^{*}q_2}{\Delta (k_1-x\Delta )}+\frac{%
q_1q_2^{*}}{\Delta ^{*}(k_1^{*}-x\Delta ^{*})}\right| ^2,\,R^{(1)}=\frac
18\left| \frac{q_1^{*}q_2}{\Delta (k_1-x\Delta )}-\frac{q_1q_2^{*}}{\Delta
^{*}(k_1^{*}-x\Delta ^{*})}\right| ^2 \,,
$$

\begin{equation}
R(++)\rightarrow R^{(2)}=\frac 12\left| \frac{(1-x)^2q_1q_2^{*}}{\Delta
^{*}(k_1-x\Delta )}+\frac{x^2q_1^{*}q_2}{\Delta (k_1^{*}-x\Delta ^{*})}%
\right| ^2\,.
\end{equation}
It allows us to regularize the integrals for $K_{gluons}^{(2)}$ in this
kinematical region.

As it is seen from above formulas for the bilinear expressions containing $%
c^{\alpha _1\alpha _2}(k_1,k_2)$, simultaneously with the infrared
singularity in the integral over $k_1$ we have the divergency at $x=0$ and $%
x=1$. The divergency at $x=1$ is related with the Regge limit $\kappa
\rightarrow \infty $. The divergency at $x=0$ has an infrared origin. The
bilinear combinations of $c^{\alpha _1\alpha _2}$ can be simplified in this
infrared region of small $k_1$ and $x$:

$$
\mid c^{+-}(k_1,k_2)\mid ^2\rightarrow \mid c^{++}(k_1,k_2)\mid
^2\rightarrow \frac{x^2\overrightarrow{q_1}^2\overrightarrow{q_2}^2}{
\overrightarrow{k_1}^2(\overrightarrow{k_1}-x\overrightarrow{\Delta })^2}\,,
$$
\begin{equation}
c^{+-}(k_1,k_2)c^{-+}(k_2,k_1)\frac{k_1^{*}}{k_1}\frac{k_2}{k_2^{*}}%
\rightarrow c^{++}(k_1,k_2)c^{++}(k_2,k_1)\frac{k_1^{*}}{k_1}\frac{k_2^{*}}{%
k_2}\rightarrow -\frac{x\overrightarrow{q_1}^2\overrightarrow{q_2}^2}{\Delta
^{*}k_1(\overrightarrow{k_1}-x\overrightarrow{\Delta })^2}\,.
\end{equation}
This singularity also can be regularized because one can calculate $R$ in
the $D$-dimensional space:

\begin{equation}
R\rightarrow \frac{x^2\overrightarrow{q_1}^2\overrightarrow{q_2}^2}{
\overrightarrow{k_1}^2(\overrightarrow{k_1}-x\overrightarrow{\Delta })^2}+
\frac{\overrightarrow{q_1}^2\overrightarrow{q_2}^2}{\overrightarrow{\Delta }%
^2(\overrightarrow{k_1}-x\overrightarrow{\Delta })^2}-\frac{x(
\overrightarrow{\Delta }\overrightarrow{k_1})\,\overrightarrow{q_1}^2
\overrightarrow{q_2}^2}{\overrightarrow{\Delta }^2\overrightarrow{k_1}^2(
\overrightarrow{k_1}-x\overrightarrow{\Delta })^2}
\end{equation}
using the approximate expression for $c^{\alpha _1\alpha _2}(k_1,k_2)$ and $%
c^{\alpha _2\alpha _1}(k_2,k_1)$ in the region of small $x$ and $k_1$:

$$
c^{\alpha _1\alpha _2}(k_1,k_2)\rightarrow x\,\frac{(\overrightarrow{k_1}%
^2\Delta ^{\alpha _1}-x\overrightarrow{\Delta }^2k_1^{\alpha _1})(
\overrightarrow{\Delta }^2q_1^{\alpha _2}-\overrightarrow{q_1}^2\Delta
^{\alpha _2})}{\overrightarrow{\Delta }^2(\overrightarrow{k_1}-x
\overrightarrow{\Delta })^2\overrightarrow{k_1}^2}\,,
$$
\begin{equation}
c^{\alpha _2\alpha _1}(k_2,k_1)\rightarrow (k_1^{\alpha _1}-x\Delta ^{\alpha
_1})\,\frac{\overrightarrow{q_2}^2\Delta ^{\alpha _2}+\overrightarrow{\Delta
}^2q_2^{\alpha _2}}{\overrightarrow{\Delta }^2(\overrightarrow{k_1}-x
\overrightarrow{\Delta })^2}.
\end{equation}
We write down also the expression for $R$ in the Regge region of small $1-x$
and fixed $\overrightarrow{k_1}$ for arbitrary $D$:
\begin{equation}
R \rightarrow \frac{\overrightarrow{q_1}^2 \overrightarrow{q_2}^2} {
\overrightarrow{k_1}^2 \overrightarrow{k_2}^2}.
\end{equation}
Let us return now to the quark-antiquark production. The total cross-section
of this process in accordance with the normalization condition for spinors
is proportional to the integral from the expression:
\begin{equation}
K_{quarks}^{(2)}=\int_0^1\frac{dx}{x(1-x)}\int d^{D-2}\overrightarrow{k_1}
\frac{8\,L}{\overrightarrow{q_1}^2\overrightarrow{q_2}^2}
\end{equation}
where we put $\delta =0$ because the integral in $x$ is convergent at $x=0$
and $x=1$. The expression for $L$ with corresponding colour factors is given
below:
\begin{equation}
L=\frac{N_c^2-1}{4N_c}\left( \frac{1-x}x\mid c(k_1,k_2)\mid ^2+\frac
x{1-x}\mid c(k_2,k_1)\mid ^2\right) +\frac
1{2N_c}Re\;c(k_1,k_2)\,c(k_2,k_1).
\end{equation}
Here two equal contributions from two different helicity states of the
quark-anti-quark pair were taken into account. Note, that in the limit of
large $N_c$ the quark contribution is smaller than gluon one and the
interference term is suppressed by the factor $1/N_c^2 $ in comparison with
the direct contribution.

The bilinear combinations of the function $c(k_1,k_2)$ entering in $L$ are
given in Appendix. In the region of  $\overrightarrow{k_1}\rightarrow x
\overrightarrow{\Delta }$ the interference for quarks
is constructive (contrary to the gluon case):

\begin{equation}
\frac 1{x^2}\left| c(k_1,k_2)\right| ^2\rightarrow \frac
1{x(1-x)}c(k_1,k_2)c(k_2,k_1)\rightarrow \left| \frac{(1-x)q_1q_2^{*}}{%
\Delta ^{*}(k_1-x\Delta )}-\frac{xq_1^{*}q_2}{\Delta (k_1^{*}-x\Delta ^{*})}%
\right| ^2.
\end{equation}
We can find the asymptotic behaviour of bilinear combinations of $c(k_1,k_2)$
at $\overrightarrow{k_1}\rightarrow x\overrightarrow{\Delta }$ for an
arbitrary dimension $D$ of space-time:

$$
\frac 1{x^2}\left| \widetilde{c}(k_1,k_2)\right| ^2\rightarrow \frac
1{x(1-x)}\widetilde{c}(k_1,k_2)\widetilde{c}(k_2,k_1)\rightarrow
$$
$$
\frac 1{32\,x(1-x)}\,\Gamma ^{+-\beta }(q_2,q_1)\Gamma ^{+-\beta ^{\prime
}}(q_2,q_1)\,\frac{tr\,\widehat{k_1}\gamma _\beta \widehat{k_2}\gamma
_{\beta ^{\prime }}}{(k_1+k_2)^4}\rightarrow
$$
\begin{equation}
\left( \frac{\overrightarrow{q_1}^2\overrightarrow{q_2}^2}{\overrightarrow{%
\Delta }^2(\overrightarrow{k_1}-x\overrightarrow{\Delta })^2}-4x(1-x)\left(
\frac{\overrightarrow{\Delta }^2\overrightarrow{q_1}-\overrightarrow{q_1}^2
\overrightarrow{\Delta }}{\overrightarrow{\Delta }^2}\,,\,\frac{
\overrightarrow{k_1}-x\overrightarrow{\Delta }}{(\overrightarrow{k_1}-x
\overrightarrow{\Delta })^2}\right) ^2\right) \frac 14tr\,(1).
\end{equation}
It gives us a possibility to regularize the infrared divergency. Futher we
use the traditional prescription $tr \,(1) \,=\,4$ for the spinor space in
the $D$-dimensional coordinate space.

\section{Infrared and collinear divergencies}

The gluon and quark production cross-sections contain infrared divergencies
which should be cancelled with the virtual corrections to the multi-Regge
processes. Using the expressions for products of amplitudes $c^{+-}$
presented in Appendix, we can calculate $R(+-)$ for $D=4$:
\begin{equation}
R(+-)=\frac{\overrightarrow{q_1}^2\overrightarrow{q_2}^2}4\left( \frac 1{
\overrightarrow{k_1}^2(\overrightarrow{k_1}-\overrightarrow{\Delta })^2}+
\frac{x^2}{\overrightarrow{k_1}^2(\overrightarrow{k_1}-x\overrightarrow{%
\Delta })^2}+\frac{(1-x)^2}{(\overrightarrow{k_1}-\overrightarrow{\Delta }%
)^2(\overrightarrow{k_1}-x\overrightarrow{\Delta })^2}\right) .
\end{equation}

As for the bilinear combinations of $c^{++}$, we write them as sums of
singular and regular terms:%
$$
\left| c^{++}(k_1,k_2)\right| ^2=\left| c^{++}(k_1,k_2)\right|
_{sing}^2+\left| c^{++}(k_1,k_2)\right| _{reg}^2\,,
$$
$$
\left| c^{++}(k_2,k_1)\right| ^2=\left| c^{++}(k_2,k_1)\right|
_{sing}^2+\left| c^{++}(k_2,k_1)\right| _{reg}^2\,,
$$
\begin{equation}
c^{++}(k_1,k_2)c^{++}(k_2,k_1)=\left( c^{++}(k_1,k_2)c^{++}(k_2,k_1)\right)
_{sing}+\left( c^{++}(k_1,k_2)c^{++}(k_2,k_1)\right) _{reg}\,,
\end{equation}
where the singular terms are chosen in accordance with the previous section
as follows:

$$
\left| c^{++}(k_1,k_2)\right| _{sing}^2=\left| c^{+-}(k_1,k_2)\right|
^2+r(k_1,x)\,,
$$

$$
Re\,\left[ \left( c^{++}(k_1,k_2)\,c^{++}(k_2,k_1)\right) _{sing}\frac{%
k_1^{*}}{k_1} \frac{k_2^{*}}{k_2}\right]=
$$
$$
=Re\,\left[ c^{+-}(k_1,k_2)\,c^{-+}(k_2,k_1)\frac{k_1^{*}}{k_1}\frac{k_2}{%
k_2^{*}} \right]-\frac 12\,\left( r(k_1,x)+r(\Delta -k_1,1-x)\right) \,,
$$
\begin{equation}
r(k_1,x)=\frac{\overrightarrow{q_1}^2\overrightarrow{q_2}^2}{\overrightarrow{
\Delta }^2}\left( x(1-x)-2\right) \frac{2x^2(1-x)\overrightarrow{k_1}
\overrightarrow{\Delta }}{\overrightarrow{k_1}^2(\overrightarrow{k_1}-x
\overrightarrow{\Delta })^2}+2\,Re\,\frac{x^3(1-x)^2q_1^2q_2^{*2}\Delta }{%
\Delta ^{*2}(k_1-x\Delta )^2k_1}\;.
\end{equation}
They are written in terms of products of the amplitudes $c^{+-}$ from
Appendix, which allows us to take into account their singularities in a
simple form. For the total sum $R_{sing}=R(+-)+R_{sing}(++)$ we obtain for $%
D=4$:

\begin{equation}
R_{sing}=2\;R(+-)+\frac 14\left( r(k_1,x)+r(\Delta -k_1,1-x)\right) .
\end{equation}
The singular contribution $R_{sing}$ contains all infrared and Regge
singularities of $R$. It decreases rapidly at large $k_1$. Integrals from
the regular terms are convergent everywhere. At fixed $x$ the quantity $R$
has singularities at $\overrightarrow{k_1}\rightarrow
0,\,\overrightarrow{k_2%
}\rightarrow 0$\thinspace and $\overrightarrow{k_1}\rightarrow x
\overrightarrow{\Delta }$. According to the results of the previous section
one can find it for an arbitrary dimension $D$ of space-time at small $k_i$
and $k_1-x\Delta $:
\begin{equation}
R\rightarrow \frac{\overrightarrow{q_1}^2\overrightarrow{q_2}^2}{
\overrightarrow{\Delta }^2\overrightarrow{k_i}^2}\,,\,R\rightarrow \frac{%
\left( 1-x(1-x)\right) ^2\overrightarrow{q_1}^2\overrightarrow{q_2}^2}{
\overrightarrow{\Delta }^2(\overrightarrow{k_1}-x\overrightarrow{\Delta })^2}
\end{equation}
after averaging over angles. The integration over these infrared regions
gives the result:
\begin{equation}
\frac{\mu ^{4-D}}\pi \int_{\inf r}d^{D-2}k_1\,R=\frac{\overrightarrow{q_1}^2
\overrightarrow{q_2}^2}{\overrightarrow{\Delta }^2}\left( \frac{\pi ^{\frac{%
D-4}2}}{\Gamma (\frac{D-2}2)}\,\frac 2{D-4}+\ln \,\frac{\Lambda ^2}{\mu ^2}%
\right) \left( 2+\left( 1-x(1-x)\right) ^2\right) ,
\end{equation}
where $\mu $ is the renormalization point and $\Lambda \ll \left| \Delta
\right| $ is an intermediate cut-off parameter: $\overrightarrow{k_i}%
^2<\Lambda ^2,\,(\overrightarrow{k_1}-x\overrightarrow{\Delta })<\Lambda ^2$%
. In the integral over the region $\overrightarrow{k_i}^2>\Lambda ^2,(
\overrightarrow{k_1}-x\overrightarrow{\Delta })>\Lambda ^2$ we can put $D=4$%
:
$$
\frac 1\pi \int d^2k_1\,R_{sing}=
$$
\begin{equation}
\frac{\overrightarrow{q_1}^2\overrightarrow{q_2}^2}{\overrightarrow{\Delta }%
^2}\left( \ln \frac{x(1-x)\overrightarrow{\Delta }^4}{\Lambda ^4}+\left(
1-x(1-x)\right) ^2\ln \,\frac{x(1-x)\overrightarrow{\Delta }^2}{\Lambda ^2}%
+x^2(1-x)^2Re\,\frac{q_1q_2^{*}}{q_1^{*}q_2}\right) .
\end{equation}
The total contribution for fixed $x$ is

$$
\frac{\mu ^{4-D}}\pi \int d^{D-2}k_1\,R_{sing}=x^2(1-x)^2Re\,\frac{%
q_1^2q_2^{*2}}{\overrightarrow{\Delta }^2}+\frac{\overrightarrow{q_1}^2
\overrightarrow{q_2}^2}{\overrightarrow{\Delta }^2}\left( 1+\left(
1-x(1-x)\right) ^2\right) \ln \left( x(1-x)\right)
$$
\begin{equation}
+\frac{\overrightarrow{q_1}^2\overrightarrow{q_2}^2}{\overrightarrow{\Delta
}%
^2}\left( \frac{\pi ^{\frac{D-4}2}}{\Gamma (\frac{D-2}2)}\,\frac 2{D-4}+\ln
\,\frac{\overrightarrow{\Delta }^2}{\mu ^2}\right) \left( 2+\left(
1-x(1-x)\right) ^2\right) .
\end{equation}
In the regions of small $x$ or $1-x$ one should substitute $R_{sing}$ by $%
2R(+-)$ and integrate it over $D-2$ dimensional transverse space:

$$
\frac{\mu ^{4-D}}\pi \int d^{D-2}k_1\,R_{sing}=
$$
\begin{equation}
\frac{\overrightarrow{q_1}^2\overrightarrow{q_2}^2}{\overrightarrow{\Delta }%
^2}\frac{2^{4-D}\pi ^{\frac{D-1}2}}{\Gamma (\frac{D-3}2)\sin \,(\pi \frac{D-
4%
}2)}\left| \frac{\overrightarrow{\Delta }^2}{\mu ^2}\right| ^{\frac{D-4}%
2}\left( 1+x^{D-4}+(1-x)^{D-4}\right) .
\end{equation}

Using the above formulas we can integrate the result also over $x$ taking
into account, that the intermediate parameter $\delta $ should be considered
as small as possible, because the virtual corrections to the amplitudes in
the multi-Regge kinematics were calculated under this condition. In
particular it means, that the contribution from the region of small $x$ and
$%
1-x$ is proportional to the expression:
$$
2\int_\delta ^\sigma \frac{d\,x}{x(1-x)}\left( 1+x^{D-4}+(1-x)^{D-4}\right)
$$
\begin{equation}
=\frac 2{D-4}+4\,\ln \,\frac \sigma \delta \,+2\ln \,\sigma \,+(D-4)\left(
\ln \,\sigma \right) ^2
\end{equation}
for small $\sigma $ and $D\rightarrow 4$. Thus, we obtain finally the
following contribution from $R_{sing}$ after its dimensional regularization:%
$$
\frac{\mu ^{4-D}}\pi \int_\delta ^{1-\delta }\frac{d\,x}{x(1-x)}\int
d^{D-2}k_1R_{sing}=\frac 16\,Re\,\frac{q_1^2q_2^{*2}}{\overrightarrow{\Delta
}^2}+\frac{\overrightarrow{q_1}^2\overrightarrow{q_2}^2}{\overrightarrow{%
\Delta }^2}\left( \frac{67}{18}-4\frac{\pi ^2}6\right)
$$
\begin{equation}
+\frac{\overrightarrow{q_1}^2\overrightarrow{q_2}^2}{\overrightarrow{\Delta
}%
^2}\frac{2^{4-D}\pi ^{\frac{D-1}2}}{\Gamma (\frac{D-3}2)\sin \,(\pi \frac{D-
4%
}2)}\left| \frac{\overrightarrow{\Delta }^2}{\mu ^2}\right| ^{\frac{D-4}%
2}\left( \frac 2{D-4}+4\ln \frac 1\delta \,-\frac{11}6\right) ,
\end{equation}
where it is implied, that the terms of the order of value of $D-4$ should be
omitted. The infrared divergencies at $D\rightarrow 4$ in the above formulas
should be cancelled with the contribution from one-loop corrections to the
Reggeon-Reggeon-particle vertex [11].

Let us consider the region of small $\Delta $:
\begin{equation}
\left| \Delta \right| \ll \left| q_1\right| \simeq \left| q_2\right|
\end{equation}
which can lead to the infrared divergency in the generalised BFKL equation
as it was in the case of LLA. Here the essential integration region
corresponds to the soft gluon transverse momenta:
\begin{equation}
k_1\sim k_2\sim \Delta \ll q
\end{equation}
where $q$ means $q_1$ or $q_2$. The expressions for $c^{+-}(k_1,k_2)$ and $%
c^{++}(k_1,k_2)$ in the soft region are given below:

$$
c^{+-}(k_1,k_2)\rightarrow c_{soft}^{+-}(k_1,k_2)=-x\frac{\overrightarrow{q}%
^2}{k_1^{*}(k_1-x\Delta )}\,,
$$

$$
c^{++}(k_1,k_2)\rightarrow c_{soft}^{++}(k_1,k_2)=
$$
\begin{equation}
=\frac{x\overrightarrow{q}^2}{\overrightarrow{\Delta }^2}\left(
\frac{k_2^2}{%
(\overrightarrow{k_1}-x\overrightarrow{\Delta })^2+x(1-x)\overrightarrow{%
\Delta }^2}-\frac{k_1}{k_1^{*}-x\Delta ^{*}}-\frac{(1-x)^2\Delta ^2}{%
(k_1-x\Delta )k_1^{*}}+\frac{(2-x)\Delta }{k_1^{*}}\right) .
\end{equation}
If we write down $c_{soft}^{++}(k_1,k_2)$ in the form:
\begin{equation}
\frac{\overrightarrow{\Delta }^2}{\overrightarrow{q}^2}%
\,c_{soft}^{++}(k_{1,}k_2)=\frac \Delta {k_1^{*}}+\chi (k_1,k_2)\,,
\end{equation}
by extracting its singularity at $\overrightarrow{k_1}\rightarrow 0$, the
following symmetry property of it can be verified:
\begin{equation}
\left( \frac \Delta {k_2^{*}}+\chi (k_2,k_1)\right) \frac{k_1^{*}}{k_1}\,
\frac{k_2^{*}}{k_2}=\frac{\Delta ^{*}}{k_2}-\chi ^{*}(k_1,k_2)\,.
\end{equation}
Using this relation we obtain for the bilinear combinations of
$c_{soft}^{++}(k_1,k_2)$ the following expressions
$$
\frac{1}{x^2\overrightarrow{q}^4}\left|
c_{soft}^{++}(k_1,k_2)\right| ^2=\left( \frac{(1-x)\left(
\overrightarrow{k_1}-x \overrightarrow{\Delta }\, , (1-
2x)\overrightarrow{k_1}
+x\overrightarrow{\Delta }\right) }{(
(\overrightarrow{k_1}-x\overrightarrow{\Delta })^2 +x(1-x)
\overrightarrow{\Delta }^2)(\overrightarrow{k_1}-x\overrightarrow{\Delta })
^2}\right) ^2
$$
$$
+\frac{(\overrightarrow{k_1}-x\overrightarrow{\Delta })
^2-(1-x)(3-4x)\overrightarrow{k_1}^2}{((\overrightarrow{k_1}-
x\overrightarrow{%
\Delta })^2+x(1-x)\overrightarrow{\Delta }^2)(\overrightarrow{k_1}-
x\overrightarrow{\Delta })^2 \overrightarrow{k_1}^2}
+\frac{1-x}{\overrightarrow{k_1}^2\left( \overrightarrow{k_1}
-x\overrightarrow{\Delta }\right) ^2}\,,
$$

$$
c_{soft}^{++}(k_1,k_2)\,c_{soft}^{++}(k_2,k_1)\,\frac{k_1^{*}}{k_1}\,\frac{%
k_2^{*}}{k_2}=
$$
\begin{equation}
=-\frac 12\left( \left| c_{soft}^{++}(k_1,k_2)\right| ^2+\left|
c_{soft}^{++}(k_2,k_1)\right| ^2\right)
+\frac{\overrightarrow{q}^2}{2k_1k_2}%
\left( c_{soft}^{++}(k_1,k_2)+c_{soft}^{++}(k_2,k_1)\right) .
\end{equation}

As it was done in a general case of fixed $\Delta $ the bilinear
combinations of $c_{soft}^{++}(k_1,k_2)$ can be presented in the following
form:%
$$
\left| c_{soft}^{++}(k_1,k_2)\right| ^2=\left| c_{soft}^{++}(k_1,k_2)\right|
_{sing}^2+\left| c_{soft}^{++}(k_1,k_2)\right| _{reg}^2\,,\,\,
c_{soft}^{++}(k_1,k_2)c_{soft}^{++}(k_2,k_1)=
$$
\begin{equation}
=\left( c_{soft}^{++}(k_1,k_2)c_{soft}^{++}(k_2,k_1)\right) _{sing}+\left(
c_{soft}^{++}(k_1,k_2)c_{soft}^{++}(k_2,k_1)\right) _{reg}
\end{equation}
where the singular soft contributions are defined as follows

$$
\left| c_{soft}^{++}(k_1,k_2)\right| _{sing}^2=\left|
c_{soft}^{+-}(k_1,k_2)\right| ^2+r_s(k_1,x)\,,
$$

$$
Re\,\left[\left( c_{soft}^{++}(k_1,k_2)\,c_{soft}^{++}(k_2,k_1)\right)
_{sing} \frac{k_1^{*}}{k_1}\frac{k_2^{*}}{k_2}\right]=
$$
\begin{equation}
=Re\,\left[ c_{soft}^{+-}(k_1,k_2)\,c_{soft}^{-+}(k_2,k_1)
\frac{k_1^{*}}{k_1%
}\frac{k_2}{k_2^{*}}\right] -\frac 12\left( r_s(k_1,x)+r_s(\Delta
-k_1,1-x)\right) \,.
\end{equation}
The bilinear combinations of amplitudes $c_{soft}^{+-}$ are obtained from
formulas of Appendix:

$$
\mid c_{soft}^{+-}(k_1,k_2)\mid ^2=\frac{x^2\overrightarrow{q}^4}{
\overrightarrow{k_1}^2(\overrightarrow{k_1}-x\overrightarrow{\Delta })^2}%
\,\,,
$$

$$
Re\,\left[c_{soft}^{+-}(k_1,k_2)\,c_{soft}^{-
+}(k_2,k_1)\,\frac{k_1^{*}}{k_1}\,
\frac{k_2}{k_2^{*}}\right] =
$$
\begin{equation}
=\frac{\overrightarrow{q}^4}2\left( \frac 1{\overrightarrow{k_1}^2(
\overrightarrow{k_1}-\overrightarrow{\Delta })^2}-
\frac{x^2}{\overrightarrow{%
k_1}^2(\overrightarrow{k_1}-x\overrightarrow{\Delta })^2}-\frac{(1-x)^2}{(
\overrightarrow{k_1}-\overrightarrow{\Delta })^2(\overrightarrow{k_1}-x
\overrightarrow{\Delta })^2}\right) .
\end{equation}
and $r_s$ is derived from $r$ at $\Delta \rightarrow 0$:

\begin{equation}
r_s(k_1,x)=\frac{\overrightarrow{q}^4}{\overrightarrow{\Delta }^2}\left(
x(1-x)-2\right) \frac{2x^2(1-x)\overrightarrow{k_1}\overrightarrow{\Delta }}{
\overrightarrow{k_1}^2(\overrightarrow{k_1}-x\overrightarrow{\Delta })^2}+2
\overrightarrow{q}^4Re\,\frac{x^3(1-x)^2\Delta }{\Delta ^{*2}(k_1-x\Delta
)^2k_1}\;.
\end{equation}

The result of integration of the singular soft terms with taking into
account the contribution of $c^{+-}$ and the dimensional regularization can
be obtained from the general case of fixed $\Delta $ by putting $\Delta
\rightarrow 0$:

$$
\frac{\mu ^{4-D}}\pi \int_\delta ^{1-\delta }\frac{d\,x}{x(1-x)}\int
d^{D-2}k_1\left( R_{soft}\right) _{sing}
$$
\begin{equation}
=\frac{\overrightarrow{q}^4}{\overrightarrow{\Delta }^2}\left( \frac{35}9-4
\frac{\pi ^2}6+\frac{2^{4-D}\pi ^{\frac{D-1}2}}{\Gamma (\frac{D-3}2)\sin
\,(\pi \frac{D-4}2)}\left| \frac{\overrightarrow{\Delta }^2}{\mu ^2}\right|
^{\frac{D-4}2}\left( \frac 2{D-4}+4\ln \frac 1\delta \,-\frac{11}6\right)
\right) \,.
\end{equation}
The regular soft terms do not contain any divergency.
Their contributions are proportional to $
\overrightarrow{q}^4/\overrightarrow{\Delta }^2$:

$$
\int_0^1\frac{d\,x}{x(1-x)}\int \frac{d^2k_1}\pi \left|
c_{soft}^{++}(k_1,k_2)\right| _{reg}^2=
$$
$$
=\frac{\overrightarrow{q}^4}{\overrightarrow{\Delta }^2}\int_0^1d\,x\,\left(
\frac 2{1-x}\,\ln \,\frac 1x\,+2\,(2-x+x^2)\ln \,\frac
x{1-x}\,+1-8x+8x^2\right) =\frac{\overrightarrow{q}^4}{\overrightarrow{%
\Delta }^2}\left( \frac{\pi ^2}3-\frac 13\right) .
$$
 For the interference
term we use the following relation which can be derived from above equations:

\begin{equation}
Re\,\left[\left( c_{soft}^{++}(k_1,k_2) c_{soft}^{++}(k_2,k_1)\right) _{reg}
\frac{k_1^*}{k_1}\frac{k_2^*}{k_2}\right]=-\frac 12\left( \left|
c_{soft}^{++}(k_1,k_2)\right| _{reg}^2+\left| c_{soft}^{++}(k_2,k_1)\right|
_{reg}^2\right)\,.
\end{equation}

Due to this relation we have
$$
\int_0^1\frac{d\,x}{x(1-x)}\int \frac{d^2k_1}\pi Re\,\left[\left(
c_{soft}^{++}(k_1,k_2)c_{soft}^{++}(k_1,k_2)\right)_{reg} \frac{k_1^{*}}{k_1}
\frac{k_2^{*}}{k_2}\right]
$$
\begin{equation}
=-\frac{\overrightarrow{q}^4}{\overrightarrow{\Delta }^2}\int_0^1d\,x\,%
\left( \frac 1{1-x}\,\ln \,\frac 1x\,+\frac 1x\,\ln \,\frac
1{1-x}+1-8x+8x^2\right) =-\frac{\overrightarrow{q}^4}{\overrightarrow{\Delta
}^2}\left( \frac{\pi ^2}3-\frac 13\right)
\end{equation}

The total contribution of the singular and regular terms in the soft region
$%
\Delta \rightarrow 0$ is

$$
\frac{\mu ^{4-D}}\pi \int_\delta ^{1-\delta }\frac{d\,x}{x(1-x)}\int
d^{D-2}k_1\,R_{soft}
$$
\begin{equation}
=\frac{\overrightarrow{q}^4}{\overrightarrow{\Delta }^2} \left(
\frac{67}{18}%
- \frac{\pi ^2}2+\frac{2^{4-D}\pi ^{\frac{D-1}2}}{\Gamma (\frac{D-3}2)\sin
\,(\pi \frac{D-4}2)}\left| \frac{\overrightarrow{\Delta }^2}{\mu ^2}\right|
^{\frac{D-4}2}\left( \frac 2{D-4}+4\ln \frac 1\delta \,-\frac{11}6\right)
\right) \,.
\end{equation}

Since the integration over $\Delta$ in the generalised BFKL equation leads
to the infrared divergency at $\Delta \,=\,0$ for $D \rightarrow 4$ , it
would be useful in the soft region $\Delta \rightarrow 0$ to obtain the
contribution of the real gluon production taking into account terms
vanishing at $D \rightarrow 4$. It can be done starting with the following
expression for the tensor $c^{\alpha_1\alpha_2}(k_1,k_2)$ (see Eq.(55)) in
the soft region:
$$
\frac{1}{x\overrightarrow{q}^2}c_{soft}^{\alpha_1\alpha_2}(k_1,k_2)=
- \frac{x k_1^{\alpha_1}((2-x)k_1^{\alpha_2}- \Delta^{\alpha_2})}
{\overrightarrow{k_1}^2(\overrightarrow{k_1}-x\overrightarrow{\Delta })^2}
$$
\begin{equation}
-(1-x)\frac{\delta^{\alpha_1\alpha_2}\left(\overrightarrow{k_1}-
x\overrightarrow{\Delta}\, , (1-2x)\overrightarrow{k_1}+
x\overrightarrow{\Delta}\right) +2x k_2^{\alpha_1}k_2^{\alpha_2}}
{2((\overrightarrow{k_1}-x\overrightarrow{\Delta })^2 +x(1-x)
\overrightarrow{\Delta }^2)(\overrightarrow{k_1}-x\overrightarrow{\Delta })
^2}\,.
\end{equation}
and
$$
\frac{1}{x\overrightarrow{q}^2}c_{soft}^{\alpha'_1\alpha_2}(k_1,k_2)
\Omega^{\alpha'_1\alpha_1}(k_1)=
- \frac{k_1^{\alpha_1}k_2^{\alpha_2}}
{(\overrightarrow{k_1}-x\overrightarrow{\Delta })^2 +x(1-x)
\overrightarrow{\Delta }^2)\overrightarrow{k_1}^2}
$$
\begin{equation}
-(1-x)\frac{\delta^{\alpha_1\alpha_2}\left(\overrightarrow{k_1}-
x\overrightarrow{\Delta}\, , (1-2x)\overrightarrow{k_1}+
x\overrightarrow{\Delta}\right) - 2(1-x)k_1^{\alpha_1}\Delta^{\alpha_2}+
2x\Delta^{\alpha_1}k_2^{\alpha_2}}
{2((\overrightarrow{k_1}-x\overrightarrow{\Delta })^2 +x(1-x)
\overrightarrow{\Delta }^2)(\overrightarrow{k_1}-x\overrightarrow{\Delta })
^2}\,.
\end{equation}
Using above equations we obtain
$$
\frac{1}{x^2\overrightarrow{q}^4}(c_{soft}^{\alpha_1\alpha_2}(k_1,k_2))^2 =
(D-2)\left( \frac{(1-x)\left(
\overrightarrow{k_1}-x \overrightarrow{\Delta }\, , (1-
2x)\overrightarrow{k_1}
+x\overrightarrow{\Delta }\right) }{2(
(\overrightarrow{k_1}-x\overrightarrow{\Delta })^2 +x(1-x)
\overrightarrow{\Delta }^2)(\overrightarrow{k_1}-x\overrightarrow{\Delta })
^2}\right) ^2
$$
\begin{equation}
+\frac{(\overrightarrow{k_1}-x\overrightarrow{\Delta })
^2-(1-x)(3-4x)\overrightarrow{k_1}^2}{2((\overrightarrow{k_1}-
x\overrightarrow{\Delta })^2+x(1-x)\overrightarrow{\Delta }^2)
(\overrightarrow{k_1}-x\overrightarrow{\Delta })^2 \overrightarrow{k_1}^2}
+\frac{2-x}{2\overrightarrow{k_1}^2( \overrightarrow{k_1}-
x\overrightarrow{\Delta }) ^2}\,,
\end{equation}
$$
c_{soft}^{\alpha_1\alpha_2}(k_1,k_2)c_{soft}^{\alpha'_2\alpha'_1}(k_2,k_1)
\Omega^{\alpha_1\alpha'_1}(k_1)\Omega^{\alpha_2\alpha'_2}(k_2)=
$$
\begin{equation}
-\frac{1}{2}\left((c_{soft}^{\alpha_1\alpha_2}(k_1,k_2))^2+
(c_{soft}^{\alpha_2\alpha_1}(k_2,k_1))^2\right)+\frac{\overrightarrow{q}^4}
{2\overrightarrow{k_1}^2\overrightarrow{k_2}^2}\,.
\end{equation}
Performing integration we obtain for arbitrary D:
$$
\frac{\mu ^{4-D}}{\pi} \int_\delta ^{1-\delta }\frac{d\,x}{x(1-x)}\int
d^{D-2}k_1\, (c_{soft}^{\alpha_1\alpha_2}(k_1,k_2))^2
=\frac{\overrightarrow{q}^4}{\overrightarrow{\Delta }^2}
\left(\frac{\pi\overrightarrow{\Delta }^2}{\mu ^2}\right)^{\frac{D-4}{2}}
\Gamma (3-\frac{D}{2})\times
$$
$$
\times\int_\delta ^{1-\delta }d\,x \left[
\frac{(2-x)\Gamma^2({D\over 2}-2)x^{D-5}}{2(1-x)\Gamma(D-4)}
+\frac{1}{2(1-x)}\int_0 ^x \frac{d\,z}{(z(1-z))^{3-{D\over 2}}}\right.
$$
$$
\left.+(x(1-x))^{{D\over 2}-2}\left({{D-2}\over 4}(1-2x)^2 +{1\over{D-
4}}\left({1\over{1-x}}-4+(6-D)x(1-x)\right)\right)\right]
$$
\begin{equation}
=\frac{\overrightarrow{q}^4}{\overrightarrow{\Delta }^2}
\left(\frac{\pi\overrightarrow{\Delta }^2}{\mu ^2}\right)^{\frac{D-4}{2}}
\Gamma (3-\frac{D}{2})\frac{\Gamma^2({D\over 2}-1)}
{\Gamma(D-2)}
\end{equation}
$$
\times\left[\frac{4}{(D-4)^2}+\frac{D-2}{2(D-4)(D-1)}+\frac{2(D-3)}
{D-4}\left( 2 \ln \frac{1}{\delta} +\psi(1)+\psi({D\over 2}-1)-2\psi(D-3)
\right)\right]\, ,
$$
where $\psi(x) =\frac{\Gamma'(x)}{\Gamma(x)}$, and
$$
\frac{\mu ^{4-D}}{\pi} \int_\delta ^{1-\delta }\frac{d\,x}{x(1-x)}\int
d^{D-2}k_1\,c_{soft}^{\alpha_1\alpha_2}(k_1,k_2)
c_{soft}^{\alpha'_2\alpha'_1}(k_2,k_1)
\Omega^{\alpha_1\alpha'_1}(k_1)\Omega^{\alpha_2\alpha'_2}(k_2)
$$
$$
=\frac{\overrightarrow{q}^4}{\overrightarrow{\Delta }^2}
\left(\frac{\pi\overrightarrow{\Delta }^2}{\mu ^2}\right)^{\frac{D-4}{2}}
\Gamma (3-\frac{D}{2})\int_\delta ^{1-\delta }d\,x
\left[-\frac{\Gamma^2({D\over 2}-2)}
{4\Gamma(D-4)}\left(\frac{2-x}{1-x}x^{D-5}+\frac{1+x}{x}(1-x)^{D-5}\right)
\right.
$$
$$
\left.+(x(1-x))^{{D\over 2}-2}\left(-{(D-2)\over 4}(1-2x)^2 +{1\over {D-
4}}\left(4-{1\over {2(1-x)}}-(6-D)x(1-x)\right)\right)\right.
$$
$$
\left.+\frac{2-x}{4x(1-x)}\int_0 ^x \frac{d\,z}{(z(1-z))^{3-{D\over 2}}}
+\frac{1+x}{4x(1-x)}\int_0 ^{1-x} \frac{d\,z}{(z(1-z))^{3-{D\over 2}}}
\right]
$$
\begin{equation}
=\frac{\overrightarrow{q}^4}{\overrightarrow{\Delta }^2}
\left(\frac{\pi\overrightarrow{\Delta }^2}{\mu ^2}\right)^{\frac{D-4}{2}}
\Gamma (3-\frac{D}{2})\frac{\Gamma^2({D\over 2}-1)}
{\Gamma(D-2)}
\end{equation}
$$
\times\left[-\frac{4}{(D-4)^2}-\frac{D-2}{2(D-4)(D-1)}+\frac{2(D-3)}
{D-4}\left( -\psi(1)-\psi({D\over 2}-1)+2\psi(D-3)
\right)\right]\, .
$$
Note, that it is possible to modify the definition of the soft and hard
parts of $c^{++}(k_1,k_2)$:
$$
c^{++}(k_1,k_2)=\widetilde{c}_{soft}^{++}(k_1,k_2)+\widetilde{c}%
_{hard}^{++}(k_1,k_2)\,,
$$
$$
\frac 1x\,\widetilde{c}_{soft}^{++}(k_1,k_2)=\frac{q_1^{*}q_2k_1(k_1-x\Delta
)-(2-x)q_1q_2^{*}\Delta k_1+q_1q_2^{*}\,\Delta ^2}{\overrightarrow{\Delta }%
^2\left( (\overrightarrow{k_1}-x\overrightarrow{\Delta })^2+x(1-x)
\overrightarrow{\Delta }^2\right) }
$$
$$
-\frac{q_1^{*}q_2k_1}{\overrightarrow{\Delta }^2(k_1^{*}-x\Delta ^{*})}%
+(2-x) \frac{q_1q_2^{*}}{\Delta ^{*}k_1^{*}}-(1-x)^2\frac{\Delta
q_1q_2^{*}}{%
\Delta ^{*}(k_1-x\Delta )k_1^{*}}\,,
$$
$$
\frac 1x\,\widetilde{c}_{hard}^{++}(k_1,k_2)=-\,\frac{Q_1^2}{(
\overrightarrow{k_1}-x\overrightarrow{q_1})^2+x(1-x)\overrightarrow{q_1}^2}
$$
\begin{equation}
+\frac{q_1^{*}k_1(k_1-x\Delta )-(2-x)q_1\Delta ^{*}k_1+q_1\overrightarrow{%
\Delta }^2}{\Delta ^{*}\left( (\overrightarrow{k_1}-x\overrightarrow{\Delta
}%
)^2+x(1-x)\overrightarrow{\Delta }^2\right) }-\frac{q_2^{*}k_1}{\Delta
^{*}k_1^{*}}
\end{equation}
in such way to include in the comparatively simple expression for $
\widetilde{c}_{soft}^{++}(k_1,k_2)$ all singular terms without the loss of
its good behaviour at large $k_1$. In the Regge limit $x\rightarrow 1$ and
fixed $k_i$ we obtain

\begin{equation}
\widetilde{c}_{soft}^{++}\rightarrow \frac{q_1q_2^{*}}{k_1^{*}k_2^{*}}
\end{equation}
and therefore in this region the contributions of the exact amplitude $%
c^{++} $ and approximate one $\widetilde{c}_{soft}^{++}$ to the differential
cross-section coincide. However, as a result of the singularity at $%
k_1=x\Delta $ after integration over $k_1$ these contributions to the total
cross-section in the Regge limit $x\rightarrow 1$ turn out to be different
(see below).

One can verify the following relation:
$$
\frac{k_1^{*}}{k_1}\,\frac{k_2^{*}}{k_2}\,\widetilde{c}%
_{soft}^{++}(k_2,k_1)= \frac{q_1q_2^{*}}{\Delta k_2}+\frac{q_1^{*}q_2}{%
\Delta k_1}+\frac{(k_1^{*}-x\Delta ^{*})(q_1q_2^{*}-q_1^{*}q_2)}{\Delta
\left( ( \overrightarrow{k_1}-x\overrightarrow{\Delta })^2+x(1-x)
\overrightarrow{\Delta }^2\right) }-\left( \widetilde{c}%
_{soft}^{++}(k_1,k_2)\right) ^{*}
$$
which can be used to simplify the interference terms after the subtraction
of the above singular terms:

$$
Re\,\left[ \left( \widetilde{c}_{soft}^{++}(k_1,k_2)\widetilde{c}%
_{soft}^{++}(k_2,k_1)\right) _{reg}\frac{k_1^{*}}{k_1}\frac{k_2^{*}}{k_2}%
\right] =-\frac 12\left( \left| \widetilde{c}_{soft}^{++}(k_1,k_2)\right|
_{reg}^2+\left| \widetilde{c}_{soft}^{++}(k_2,k_1)\right| _{reg}^2\right)
$$
$$
+Re\,\left[ \frac{q_1q_2^{*}-q_1^{*}q_2}{2\Delta }\,\left( \frac{k_1-k_2}{%
2k_1k_2}+\frac{k_1^{*}-x\Delta ^{*}}{(\overrightarrow{k_1}-x\overrightarrow{%
\Delta })^2+x(1-x)\overrightarrow{\Delta }^2}\right) \left( \widetilde{c}%
_{soft}^{++}(k_1,k_2)-\widetilde{c}_{soft}^{++}(k_2,k_1)\right) \right]
$$
\begin{equation}
-\frac{\overrightarrow{q_1}^2\overrightarrow{q_2}^2}{2\,\overrightarrow{k_1}%
^2\overrightarrow{k_2}^2}+Re\,\left[ \frac{\overrightarrow{q_1}
\overrightarrow{q_2}}{2k_1k_2}\,\left( \widetilde{c}_{soft}^{++}(k_1,k_2)+
\widetilde{c}_{soft}^{++}(k_2,k_1)\right) \right] .
\end{equation}
The result of the integration of this contribution is%
$$
\int_0^{1-\widetilde{\delta }}\frac{d\,x}{x(1-x)}\int \frac{d^2k_1}\pi
\left| \widetilde{c}_{soft}^{++}(k_1,k_2)\right| _{reg}^2=
$$
$$
=\frac{\overrightarrow{q_1}^2\overrightarrow{q_2}^2}{\overrightarrow{\Delta
}%
^2}\int_0^1d\,x\,\left( \frac 2{1-x}\,\ln \,\frac 1x\,+2\,(2-x+x^2)\ln
\,\frac x{1-x}\,+1-8x+8x^2\right)
$$
\begin{equation}
+\frac 4{\overrightarrow{\Delta }^2}\left( \overrightarrow{q_1}^2
\overrightarrow{q_2}^2-(\overrightarrow{q_1}\overrightarrow{q_2})^2\right)
\int_0^{1-\widetilde{\delta }}d\,x\,\frac x{1-x}(1-4x+2x^2)\,,
\end{equation}
$$
\int_{\widetilde{\delta }}^{1-\widetilde{\delta }}\frac{d\,x}{x(1-x)}\int
\frac{d^2k_1}\pi \,Re\,\left[ \left( \widetilde{c}_{soft}^{++}(k_1,k_2)
\widetilde{c}_{soft}^{++}(k_2,k_1)\right) _{reg}\frac{k_1^{*}}{k_1}\frac{%
k_2^{*}}{k_2}\right]=
$$
\begin{equation}
-\int_{\widetilde{\delta }%
}^{1-\widetilde{\delta }}\frac{d\,x}{x(1-x)}
\left[ 2\,\frac{\overrightarrow{q_1}^2\overrightarrow{q_2}^2-
(\overrightarrow{q_1}\overrightarrow{q_2})^2}{\overrightarrow{\Delta}^2}
+ \int
\frac{d^2k_1}\pi \,\frac 12\left( \left| \widetilde{c}_{soft}^{++}(k_1,k_2)%
\right| _{reg}^2+\left| \widetilde{c}_{soft}^{++}(k_2,k_1)\right|
_{reg}^2\right)\right],
\end{equation}
where we introduced the infinitesimal parameter $\widetilde{\delta }$
different from $\delta $ to show, that the divergency at $x=1$ of the
regularized bilinear combinations of $\widetilde{c}$ should cancel in the
total regularized contribution of $c$. This cancellation will be
demonstrated in the next paper.

We return now to the quark production amplitude $c(k_1,k_2)$ and present it
in the form analogous to the gluon case:
\begin{equation}
c(k_1,k_2)=c_{sing}(k_1,k_2)+c_{reg}(k_1,k_2)\,,
\end{equation}
where the singular and regular terms are chosen as follows

$$
\frac 1xc_{sing}(k_1,k_2)=-\frac{(1-x)\Delta q_1q_2^{*}k_1^{*}-x\Delta
^{*}q_1^{*}q_2k_1+x\overrightarrow{\Delta }^2q_1^{*}q_2}{\overrightarrow{%
\Delta }^2\left( (\overrightarrow{k_1}-x\overrightarrow{\Delta })^2+x(1-x)
\overrightarrow{\Delta }^2\right) }+\frac{(1-x)q_1q_2^{*}}{\Delta
^{*}(k_1-x\Delta )}-\frac{xq_1^{*}q_2}{\Delta (k_1^{*}-x\Delta ^{*})}
$$
and

\begin{equation}
\frac 1xc_{reg}(k_1,k_2)=\frac{(1-x)q_1k_1^{*}-xq_1^{*}k_1+x\overrightarrow{%
q_1}^2}{\left( (\overrightarrow{k_1}-x\overrightarrow{q_1})^2+x(1-x)
\overrightarrow{q_1}^2\right) }-\frac{(1-x)q_1k_1^{*}-xq_1^{*}k_1+xq_1^{*}%
\Delta }{\left( (\overrightarrow{k_1}-x\overrightarrow{\Delta })^2+x(1-x)
\overrightarrow{\Delta }^2\right) }\,.
\end{equation}
The term $c_{reg}$ does not lead to any divergency and give a regular
contribution to the BFKL kernel at the soft region $\Delta \rightarrow 0$.
The singular term $c_{sing}$ has the important symmetry property:
\begin{equation}
\frac 1{1-x}\,c_{sing}(k_2,k_1)=\frac 1x\,c_{sing}^{*}(k_1,k_2)\,,
\end{equation}
which is obvious from its another representation:
$$
\frac 1xc_{sing}(k_1,k_2)=
$$
\begin{equation}
=x(1-x)\,\frac{(1-x)\Delta q_1q_2^{*}(k_1^{*}-x\Delta ^{*})-x\Delta
^{*}q_1^{*}q_2(k_1-x\Delta )-2\,(\overrightarrow{k_1}-x\overrightarrow{%
\Delta })^2\overrightarrow{q_1}\overrightarrow{q_2}}{(\overrightarrow{k_1}-x
\overrightarrow{\Delta })^2\left( (\overrightarrow{k_1}-x\overrightarrow{%
\Delta })^2+x(1-x)\overrightarrow{\Delta }^2\right) }\,.
\end{equation}
Using this property, we obtain the following contributions for the bilinear
combinations of $c_{sing}$ taking into account the dimensional
regularization of the singularity at $\overrightarrow{k_1}\rightarrow x
\overrightarrow{\Delta }$ (using eq. (111) and averaging it over angles):

$$
\frac{\mu ^{4-D}}{\pi}\int_0^1\frac{d\,x}{x^2}\int d^{D-2}\,k_1
\left|c_{sing}(k_1,k_2)\right| ^2=\frac{\mu ^{4-D}}{\pi}\int_0^1
\frac{d\,x}{x(1-x)}\int d^{D-2}\,k_1 \,c_{sing}(k_1,k_2)c_{sing}(k_2,k_1)\,
$$
$$
=\frac{\mu ^{4-D}}{\pi}\int_0^1 d\,x\,x^2(1-x)^2\int d^{D-2}\,k_1,\frac{
\overrightarrow{\Delta }^2\overrightarrow{q_1}^2\overrightarrow{q_2}^2\left(
1-\frac{4x(1-x)}{D-2}\right) +4(\overrightarrow{k_1}-x\overrightarrow{\Delta
})^2\left( \overrightarrow{q_1},\overrightarrow{q_2}\right) ^2}{(
\overrightarrow{k_1}-x\overrightarrow{\Delta })^2\left(
(\overrightarrow{k_1}%
-x\overrightarrow{\Delta })^2+x(1-x)\overrightarrow{\Delta }^2\right) ^2}
$$
$$
=\frac{\Gamma (3-\frac D2)}{\overrightarrow{\Delta }^2}\int_0^1d\,x\,
\left( \frac{\pi \overrightarrow{\Delta }^2}{\mu ^2}x(1-x)\right) ^{\frac
D2-2}\times
$$
$$
\times\left( 4x(1-x)(\overrightarrow{q_1},\overrightarrow{q_2})^2+\frac{6-
D}{%
D-4}\left( 1-\frac{4x(1-x)}{D-2}\right) \overrightarrow{q_1}^2
\overrightarrow{q_2}^2\right)
$$
\begin{equation}
=\frac 1{\overrightarrow{\Delta }^2}\left( \frac{\pi \overrightarrow{\Delta
}%
^2}{\mu ^2}\right) ^{\frac D2-2}\Gamma (3-\frac D2)\frac{\Gamma (\frac
D2)\Gamma (\frac D2)}{\Gamma (D)}\,4\,\left( \frac{6-D}{D-4}\,
\overrightarrow{q_1}^2\overrightarrow{q_2}^2+\left( \overrightarrow{q_1},
\overrightarrow{q_2}\right) ^2\right)
\end{equation}

At fixed $\overrightarrow{\Delta }$ the singularity of this expression at $%
D\rightarrow 4$ is cancelled with the contribution from the fermion
correction to the RRP vertex $\left[ 12\right] $. The integration over $%
\Delta $ in the generalised BFKL equation leads to the infrared divergency
at $\Delta =0$ for $D\rightarrow 4$:

$$
-16N_cn_f\pi ^{\frac D2-1}\left( \frac{g^2\,\mu ^{4-D}}{2(2\pi )^{D-1}}%
\right) ^2\Gamma (2-\frac D2)\frac{\Gamma (\frac D2)\Gamma (\frac D2)}{%
\Gamma (D)}\int d^{D-2}\Delta \left( \overrightarrow{\Delta }^2\right)
^{\frac D2-3}.
$$
The quark correction to the RRP vertex expressed in terms of the bare
coupling constant does not give the singular contribution at small $\Delta $%
. This divergency is cancelled with the doubled contribution of the quark
correction to the gluon Regge trajectory $\left[ 13\right] $:

$$
\omega _q^{(2)}(-\overrightarrow{q}^2)=\frac{N_cn_f\pi ^{\frac D2-1}g^4}{%
(2\pi )^{2D-2}\mu ^{D-4}}\,\Gamma (2-\frac D2)\frac{\Gamma ^2(\frac D2)}{%
\Gamma (D)}\int \frac{d^{D-2}q_1\,\overrightarrow{q}^2}{\overrightarrow{q_1}%
^2\left( \overrightarrow{q}-\overrightarrow{q_1}\right) ^2}\left( 2(\frac{
\overrightarrow{q_1}}\mu )^{D-4}-(\frac{\overrightarrow{q}}\mu
)^{D-4}\right) .
$$

\section{Conclusion}

In this paper using the helicity representation we simplified the gluon and
quark production amplitudes in the quasi-multi-Regge kinematics for the
final states (see eqs (59) and (72)). The corresponding next to leading
contributions to the integral kernel of the BFKL equation were expressed in
terms of the integrals from the squares of these helicity amplitudes over
transverse and longitudinal momenta of produced particles (see eqs (82) and
(108)). All infrared divergencies are extracted from these expressions in an
explicit form and are regularized in the $D$-dimensional space. These
divergencies should cancel with the analogous divergencies from the virtual
corrections to the BFKL equation which were calculated earlier in
refs. [11]-[13]. In the end of the previous section the cancellation of the
infrared divergencies was demonstrated for the fermion contribution.
The total
next to leading correction to the integral kernel can be calculated in an
explicit form in terms of the dilogarithm integrals. We shall do it in our
next publications.

\vspace{1cm} \noindent

{\large {\bf Acknowledgements}}\\

We want to thank Universit\"at Hamburg, DESY-Hamburg and DESY-Zeuthen for
the hospitality during our stay in Germany. One of us (L.N.L.) is indebted
to the Alexander-von-Humboldt Foundation for the award, which gave him a
possibility to work on this problem during the last year. The fruitful
discussions with J. Bartels, J. Bl\"umlein, V. Del Duca, E. Kuraev, H.
Lotter and T. Ohl were very helpful. \newpage
\noindent

{\Large {\bf Appendix}}

For quantities entering in $R(+-)$ we obtain the simple expressions:
$$
\mid c^{+-}(k_1,k_2)\mid ^2=\frac{\overrightarrow{q_1}^2\overrightarrow{q_2}%
^2}{\overrightarrow{\Delta }^2}\mid \frac 1{k_1-x\Delta }-\frac 1{k_1}\mid
^2=\frac{x^2\overrightarrow{q_1}^2\overrightarrow{q_2}^2}{\overrightarrow{k_1
}^2(\overrightarrow{k_1}-x\overrightarrow{\Delta })^2}\,\,,
$$

$$
Re\,c^{+-}(k_1,k_2)\,c^{-
+}(k_2,k_1)\,\frac{k_1^{*}}{k_1}\,\frac{k_2}{k_2^{*}%
}=
$$
$$
=\frac{\overrightarrow{q_1}^2\overrightarrow{q_2}^2}{\overrightarrow{\Delta
}%
^2}Re\,\left( \frac 1{k_1-x\Delta }-\frac 1{k_1}\right) \left( \frac
1{x\,\Delta ^{*}-k_1^{*}}-\frac 1{\Delta ^{*}-k_1^{*}}\right)
$$
$$
=\frac{\overrightarrow{q_1}^2\overrightarrow{q_2}^2}2\left( \frac 1{
\overrightarrow{k_1}^2(\overrightarrow{k_1}-\overrightarrow{\Delta })^2}-
\frac{x^2}{\overrightarrow{k_1}^2(\overrightarrow{k_1}-x\overrightarrow{%
\Delta })^2}-\frac{(1-x)^2}{(\overrightarrow{k_1}-\overrightarrow{\Delta }%
)^2(\overrightarrow{k_1}-x\overrightarrow{\Delta })^2}\right) .
$$

More complicated expressions are obtained for the second contribution. We
begin with the squared amplitude:

$$
x^{-2}\,\mid c^{++}(k_1,k_2)\mid ^2=4\frac{\overrightarrow{q_1}^2}{
\overrightarrow{\Delta }^2}\,\frac{[(\overrightarrow{q_1}-
\overrightarrow{k_1%
})\times (\overrightarrow{\Delta }-\overrightarrow{k_1})]^2}{(
\overrightarrow{k_1}^2-2x\overrightarrow{q_1}\overrightarrow{k_1}+x
\overrightarrow{q_1}^2)(\overrightarrow{k_1}^2-2x\overrightarrow{\Delta }
\overrightarrow{k_1}+x\overrightarrow{\Delta }^2)}
$$
$$
+\left[ 1-\frac{\overrightarrow{q_1}^2}{\overrightarrow{\Delta }^2}%
+(1-x)\left( \frac{\overrightarrow{q_1}^2-2\overrightarrow{q_1}
\overrightarrow{k_1}}{\overrightarrow{k_1}^2-2x\overrightarrow{q_1}
\overrightarrow{k_1}+x\overrightarrow{q_1}^2}-\frac{\overrightarrow{q_1}^2}{
\overrightarrow{\Delta }^2}\frac{\overrightarrow{\Delta }^2-
2\overrightarrow{%
\Delta }\overrightarrow{k_1}}{\overrightarrow{k_1}^2-2x\overrightarrow{%
\Delta }\overrightarrow{k_1}+x\overrightarrow{\Delta }^2}\right) \right] ^2
$$
$$
+2\,Re\,\left[ \left( \frac{(q_1-k_1)^2}{\overrightarrow{k_1}^2-2x
\overrightarrow{q_1}\overrightarrow{k_1}+x\overrightarrow{q_1}^2}-\frac{
\overrightarrow{q_1}^2}{\overrightarrow{\Delta }^2}\frac{(\Delta -k_1)^2}{
\overrightarrow{k_1}^2-2x\overrightarrow{\Delta } \overrightarrow{k_1}+x
\overrightarrow{\Delta }^2}\right) \right.
$$
$$
\left. \left( \frac{q_2}{\Delta k_1}\left( \frac{(1-x)^2\Delta ^{*}q_1^{*}}{%
k_1^{*}-x\Delta ^{*}}-q_1^{*}(2-x)+k_1^{*}\right) +\frac{k_1^{*}q_1q_2^{*}}{
\overrightarrow{\Delta }^2(k_1-x\Delta )}\right) \right]
$$
$$
+\frac{(1-x)^2}{x^2}\,\frac{\overrightarrow{q_1}^2\overrightarrow{q_2}^2}{
\overrightarrow{\Delta }^2}\mid \frac{1-x}{k_1-x\Delta }-\frac 1{k_1}\mid
^2+ \frac{\overrightarrow{q_1}^2\overrightarrow{q_2}^2}{\overrightarrow{%
\Delta }^4}\,\mid 1+\frac{x\Delta }{k_1-x\Delta }\mid ^2
$$
$$
+\frac{\overrightarrow{q_2}^2}{\overrightarrow{\Delta }^2}\mid \frac{q_1 }{%
k_1}-1\mid ^2+\frac{1-x}{\overrightarrow{\Delta }^2}\,2\,Re\,\,\frac{%
q_1^2(q_2^{*})^2}{\Delta ^{*}}\left( \frac{(1-x)\Delta }{(k_1-x\Delta )^2}%
-\frac 1{k_1-x\Delta }\right)
$$
$$
-\frac{2\overrightarrow{q_2}^2}{\overrightarrow{\Delta }^2}Re\left[ \frac{1-
x%
}xq_1\left( \frac{1-x}{k_1-x\Delta }-\frac 1{k_1}\right) \left( \frac{%
q_1^{*} }{k_1^{*}}-1\right) +\frac{q_1^{*}q_2}{\Delta q_2^{*}}\left( \frac{%
q_1^{*}-x\Delta ^{*}}{k_1^{*}-x\Delta ^{*}}-1\right) \right] .
$$
It can be presented in the following form convenient for the subsequent
integration:

$$
x^{-2}\mid c^{++}(k_1,k_2)\mid ^2=
$$
$$
=(1-x)^2\left[ \frac{\overrightarrow{q_1}^2-2\overrightarrow{k_1}
\overrightarrow{q_1}}{(\overrightarrow{k_1}-x\overrightarrow{q_1})^2+x(1-x)
\overrightarrow{q_1}^2}-\frac{\overrightarrow{q_1}^2}{\overrightarrow{\Delta
}^2}\,\frac{\overrightarrow{\Delta }^2-2\overrightarrow{k_1}\overrightarrow{%
\Delta }}{(\overrightarrow{k_1}-x\overrightarrow{\Delta })^2+x(1-x)
\overrightarrow{\Delta }^2}\right] ^2
$$
$$
-2(1-x)\overrightarrow{q_1}^2\,\frac{(\overrightarrow{q_2},\overrightarrow{%
q_1}-\overrightarrow{k_1})^2+(\overrightarrow{q_2},\overrightarrow{\Delta }-
\overrightarrow{k_1})^2-\overrightarrow{q_2}^2(\overrightarrow{q_2}^2+
\overrightarrow{q_1}\overrightarrow{\Delta }-\overrightarrow{k_1}(
\overrightarrow{q_1}+\overrightarrow{\Delta }))}{\overrightarrow{\Delta }%
^2\left( (\overrightarrow{k_1}-x\overrightarrow{q_1})^2+x(1-x)
\overrightarrow{q_1}^2\right) \left( (\overrightarrow{k_1}-x\overrightarrow{%
\Delta })^2+x(1-x)\overrightarrow{\Delta }^2\right) }
$$

$$
+2\overrightarrow{q_1}^2\overrightarrow{q_2}^2\,\frac{(\overrightarrow{q_1}-
\overrightarrow{k_1},\overrightarrow{\Delta }-\overrightarrow{k_1})}{
\overrightarrow{\Delta }^2\left( (\overrightarrow{k_1}-x\overrightarrow{q_1}%
)^2+x(1-x)\overrightarrow{q_1}^2\right) \left( (\overrightarrow{k_1}-x
\overrightarrow{\Delta })^2+x(1-x)\overrightarrow{\Delta }^2\right) }
$$
$$
+\frac{\overrightarrow{q_1}^2\overrightarrow{q_2}^2}{\overrightarrow{\Delta
}%
^2}\left( \frac{2-x}{x\overrightarrow{k_1}^2}-\frac{2-x}{x(\overrightarrow{%
k_1}-x\overrightarrow{\Delta })^2}+\frac{4-4x+2x^2}{(\overrightarrow{k_1}-x
\overrightarrow{\Delta })^2}-\frac{4-4x+2x^2}{(\overrightarrow{k_1}-x
\overrightarrow{\Delta })^2+x(1-x)\overrightarrow{\Delta }^2}\right)
$$
$$
+\frac{(1-x)^2\overrightarrow{q_1}^2\overrightarrow{q_2}^2}{\overrightarrow{%
k_1}^2(\overrightarrow{k_1}-x\overrightarrow{\Delta })^2}-\frac{2(1-x+x^2)
\overrightarrow{q_1}^2\overrightarrow{q_2}^2}{\left( (\overrightarrow{k_1}-x
\overrightarrow{q_1})^2+x(1-x)\overrightarrow{q_1}^2\right) \overrightarrow{%
\Delta }^2}
$$
$$
+\frac{2(1-x)\overrightarrow{q_1}^2}{\overrightarrow{\Delta }^2}\left(
\frac{%
(3-2x)\overrightarrow{q_1}\overrightarrow{q_2}-(2-x)\overrightarrow{q_2}^2}{%
( \overrightarrow{k_1}-x\overrightarrow{q_1})^2+x(1-
x)\overrightarrow{q_1}^2}%
- \frac{(3-2x)\overrightarrow{q_1}\overrightarrow{q_2}-(2-x)\overrightarrow{%
q_2}^2}{(\overrightarrow{k_1}-x\overrightarrow{\Delta })^2+x(1-x)
\overrightarrow{\Delta }^2}\right)
$$
$$
+2\,Re\,\left( (1-x)\frac{(q_1)^2}{\overrightarrow{\Delta }^2}\,\frac{%
q_1^{*}(x\Delta q_2^{*}-2\Delta ^{*}q_2)+k_1^{*}(\Delta ^{*}q_2-x\Delta
q_2^{*})}{k_1\left( (\overrightarrow{k_1}-x\overrightarrow{q_1})^2+x(1-x)
\overrightarrow{q_1}^2\right) }\right.
$$
$$
\left. +\frac{q_1q_2^{*}}{\overrightarrow{\Delta }^2}\,\frac{%
xq_1^{*}(q_1-x\Delta )^2-2(1-x)^2q_1^{*}\Delta
^2+(1-x)k_1^{*}q_1(q_1-x\Delta )+(1-x)^2k_1^{*}\Delta ^2}{(k_1-x\Delta
)\left( (\overrightarrow{k_1}-x\overrightarrow{\Delta })^2+x(1-x)
\overrightarrow{\Delta }^2\right) }\right.
$$
$$
\left. +\frac{x^2(1-x)^2\overrightarrow{q_1}^2q_1q_2\Delta ^{*}}{\Delta
k_1(k_1^{*}-x\Delta ^{*})\left( (\overrightarrow{k_1}-x\overrightarrow{q_1}%
)^2+x(1-x)\overrightarrow{q_1}^2\right) }\right.
$$
$$
\left. +\frac{x^2\overrightarrow{q_1}^2q_2\Delta \left(
2(1-x)q_1^{*}+xq_2^{*}-(1-x)k_1^{*}\right) }{\overrightarrow{\Delta }%
^2k_1\left( (\overrightarrow{k_1}-x\overrightarrow{\Delta })^2+x(1-x)
\overrightarrow{\Delta }^2\right) }\right.
$$
$$
\left. +\frac{x^2(1-x)^2\overrightarrow{q_1}^2q_1^{*}q_2\left(
2(1-x)k_1\Delta -2k_1^2-\Delta ^2\right) }{\Delta ^2 k_1(k_1^{*}-x\Delta
^{*})\left( ( \overrightarrow{k_1}-x\overrightarrow{\Delta })^2+x(1-x)
\overrightarrow{\Delta }^2\right) }\right.
$$
$$
\left. +\frac{x^2(1-x)q_1q_2^{*}\Delta ((1-x)q_2+xq_1)}{\overrightarrow{%
\Delta }^2k_1(k_1-x\Delta )}+\frac{x^2(1-x)^2(q_1^{*}q_2)^2}{\Delta
^2(k_1^{*}-x\Delta ^{*})^2}\,\right) .
$$
For the interference term we obtain even more complicated expression:
$$
c^{++}(k_1,k_2)\,c^{++}(k_2,k_1)\,\frac{k_1^{*}}{k_1}\,\frac{k_2^{*}}{k_2}=
$$
$$
\frac{x(1-x)q_1^4\left( x(2-x)q_1^{*}+(1-x)^2k_1^{*}\right) \left(
q_1^{*}-x^2(q_2^{*}+k_1^{*})\right) }{k_1(\Delta -k_1)\left( (
\overrightarrow{k_1}-x\overrightarrow{q_1})^2+x(1-x)\overrightarrow{q_1}%
^2\right) \left( (\overrightarrow{k_1}+\overrightarrow{q_2}-
x\overrightarrow{%
q_1})^2+x(1-x)\overrightarrow{q_1}^2\right) }
$$
$$
-\frac{x(1-x)^2q_1\overrightarrow{q_1}^2\left(
x(2-x)q_1^{*}+(1-x)^2k_1^{*}\right) \left( (2+x)q_1-\Delta +k_1\right) }{%
k_1\left( (\overrightarrow{k_1}-x\overrightarrow{q_1})^2+x(1-x)
\overrightarrow{q_1}^2\right) \left( (\overrightarrow{k_1}+\overrightarrow{%
q_2}-x\overrightarrow{q_1})^2+x(1-x)\overrightarrow{q_1}^2\right) }
$$
$$
-\frac{x^2(1-x)q_1\overrightarrow{q_1}^2\left( (3-x)q_1-k_1\right) \left(
q_1^{*}-x^2(q_2^{*}+k_1^{*})\right) }{(\Delta -k_1)\left( (\overrightarrow{%
k_1}-x\overrightarrow{q_1})^2+x(1-x)\overrightarrow{q_1}^2\right) \left( (
\overrightarrow{k_1}+\overrightarrow{q_2}-x\overrightarrow{q_1})^2+x(1-x)
\overrightarrow{q_1}^2\right) }
$$
$$
+\frac{x^2(1-x)^2(q_1^{*})^2\left( (3-x)q_1-k_1\right) \left(
(2+x)q_1-\Delta +k_1\right) }{\left( (\overrightarrow{k_1}-x\overrightarrow{%
q_1})^2+x(1-x)\overrightarrow{q_1}^2\right) \left( (\overrightarrow{k_1}+
\overrightarrow{q_2}-x\overrightarrow{q_1})^2+x(1-x)\overrightarrow{q_1}%
^2\right) }
$$
$$
+\frac{x(1-x)\overrightarrow{q_1}^4\Delta ^2\left( x(2-x)\Delta
^{*}+(1-x)^2k_1^{*}\right) \left( \Delta ^{*}-x^2k_1^{*}\right) }{(\Delta
^{*})^2k_1(\Delta -k_1)\left( (\overrightarrow{k_1}-x\overrightarrow{\Delta
}%
)^2+x(1-x)\overrightarrow{\Delta }^2\right) ^2}
$$
$$
-\frac{x(1-x)^2\overrightarrow{q_1}^4\left( x(2-x)\Delta
^{*}+(1-x)^2k_1^{*}\right) \left( (1+x)\Delta +k_1\right) }{k_1\Delta
^{*}\left( (\overrightarrow{k_1}-x\overrightarrow{\Delta })^2+x(1-x)
\overrightarrow{\Delta }^2\right) ^2}
$$
$$
-\frac{x^2(1-x)\overrightarrow{q_1}^4\left( (3-x)\Delta -k_1\right) \left(
\Delta ^{*}-x^2k_1^{*}\right) }{\Delta ^{*}(\Delta -k_1)\left( (
\overrightarrow{k_1}-x\overrightarrow{\Delta })^2+x(1-x)\overrightarrow{%
\Delta }^2\right) ^2}
$$
$$
+\frac{x^2(1-x)^2\overrightarrow{q_1}^4\left( (3-x)\Delta -k_1\right) \left(
(1+x)\Delta +k_1\right) }{\Delta ^2\left( (\overrightarrow{k_1}-x
\overrightarrow{\Delta })^2+x(1-x)\overrightarrow{\Delta }^2\right) ^2}
$$
$$
-\frac{x(1-x)q_1^2\overrightarrow{q_1}^2\Delta \left(
x(2-x)q_1^{*}+(1-x)^2k_1^{*}\right) \left( \Delta ^{*}-x^2k_1^{*}\right) }{%
k_1\Delta ^{*}(\Delta -k_1)\left( (\overrightarrow{k_1}-
x\overrightarrow{q_1}%
)^2+x(1-x)\overrightarrow{q_1}^2\right) \left( (\overrightarrow{k_1}-x
\overrightarrow{\Delta })^2+x(1-x)\overrightarrow{\Delta }^2\right) }
$$
$$
+\frac{x(1-x)^2q_1^2\overrightarrow{q_1}^2\left(
x(2-x)q_1^{*}+(1-x)^2k_1^{*}\right) \left( (1+x)\Delta +k_1\right) }{%
k_1\Delta \left( (\overrightarrow{k_1}-x\overrightarrow{q_1})^2+x(1-x)
\overrightarrow{q_1}^2\right) \left( (\overrightarrow{k_1}-x\overrightarrow{%
\Delta })^2+x(1-x)\overrightarrow{\Delta }^2\right) }
$$
$$
+\frac{x^2(1-x)q_1^{*}\overrightarrow{q_1}^2\Delta \left(
(3-x)q_1-k_1\right) \left( \Delta ^{*}-x^2k_1^{*}\right) }{\Delta
^{*}(\Delta -k_1)\left( (\overrightarrow{k_1}-x\overrightarrow{q_1}%
)^2+x(1-x) \overrightarrow{q_1}^2\right) \left( (\overrightarrow{k_1}-x
\overrightarrow{\Delta })^2+x(1-x)\overrightarrow{\Delta }^2\right) }
$$
$$
-\frac{x^2(1-x)^2q_1^{*}\overrightarrow{q_1}^2\left( (3-x)q_1-k_1\right)
\left( (1+x)\Delta +k_1\right) }{\Delta \left( (\overrightarrow{k_1}-x
\overrightarrow{q_1})^2+x(1-x)\overrightarrow{q_1}^2\right) \left( (
\overrightarrow{k_1}-x\overrightarrow{\Delta })^2+x(1-x)\overrightarrow{%
\Delta }^2\right) }
$$
$$
-\frac{x(1-x)q_1^2\overrightarrow{q_1}^2\Delta \left( x(2-x)\Delta
^{*}+(1-x)^2k_1^{*}\right) \left( q_1^{*}-x^2(q_2^{*}+k_1^{*})\right) }{%
\Delta ^{*}k_1(\Delta -k_1)\left( (\overrightarrow{k_1}-x\overrightarrow{%
\Delta })^2+x(1-x)\overrightarrow{\Delta }^2\right) \left( (\overrightarrow{%
k_1}+\overrightarrow{q_2}-x\overrightarrow{q_1})^2+x(1-
x)\overrightarrow{q_1}%
^2\right) }
$$
$$
+\frac{x(1-x)^2q_1^{*}\overrightarrow{q_1}^2\Delta \left( x(2-x)\Delta
^{*}+(1-x)^2k_1^{*}\right) \left( (2+x)q_1-\Delta +k_1\right) }{\Delta
^{*}k_1\left( (\overrightarrow{k_1}-x\overrightarrow{\Delta })^2+x(1-x)
\overrightarrow{\Delta }^2\right) \left( (\overrightarrow{k_1}+
\overrightarrow{q_2}-x\overrightarrow{q_1})^2+x(1-x)\overrightarrow{q_1}%
^2\right) }
$$
$$
+\frac{x^2(1-x)q_1^2\overrightarrow{q_1}^2\left( (3-x)\Delta -k_1\right)
\left( q_1^{*}-x^2(q_2^{*}+k_1^{*})\right) }{\Delta (\Delta -k_1)\left( (
\overrightarrow{k_1}-x\overrightarrow{\Delta })^2+x(1-x)\overrightarrow{%
\Delta }^2\right) \left( (\overrightarrow{k_1}+\overrightarrow{q_2}-x
\overrightarrow{q_1})^2+x(1-x)\overrightarrow{q_1}^2\right) }
$$
$$
-\frac{x^2(1-x)^2q_1^{*}\overrightarrow{q_1}^2\left( (3-x)\Delta -k_1\right)
\left( (2+x)q_1-\Delta +k_1\right) }{\Delta \left( (\overrightarrow{k_1}-x
\overrightarrow{\Delta })^2+x(1-x)\overrightarrow{\Delta }^2\right) \left( (
\overrightarrow{k_1}+\overrightarrow{q_2}-x\overrightarrow{q_1})^2+x(1-x)
\overrightarrow{q_1}^2\right) }
$$
$$
+x^3q_1q_2^{*}\,\frac{xq_1^{*}k_1\left( (3-x)q_1-k_1\right) -q_1^2\left(
x(2-x)q_1^{*}+(1-x)^2k_1^{*}\right) }{\Delta ^{*}k_1(k_1-x\Delta )\left( (
\overrightarrow{k_1}-x\overrightarrow{q_1})^2+x(1-x)\overrightarrow{q_1}%
^2\right) }
$$
$$
-x^3q_1q_2^{*}\,\frac{xq_1^{*}k_1\left( (3-x)q_1-k_1\right) -q_1^2\left(
x(2-x)q_1^{*}+(1-x)^2k_1^{*}\right) }{\Delta ^{*}k_1(k_1-\Delta )\left( (
\overrightarrow{k_1}-x\overrightarrow{q_1})^2+x(1-x)\overrightarrow{q_1}%
^2\right) }
$$
$$
+x(1-x)^2q_1^{*}q_2\,\frac{xq_1^{*}k_1\left( (3-x)q_1-k_1\right)
-q_1^2\left( x(2-x)q_1^{*}+(1-x)^2k_1^{*}\right) }{\Delta
k_1(k_1^{*}-x\Delta ^{*})\left( (\overrightarrow{k_1}-x\overrightarrow{q_1}%
)^2+x(1-x)\overrightarrow{q_1}^2\right) }
$$
$$
+x^3\overrightarrow{q_1}^2q_1q_2^{*}\,\frac{x\Delta ^{*}k_1\left(
(x-3)\Delta +k_1\right) +\Delta ^2\left( x(2-x)\Delta
^{*}+(1-x)^2k_1^{*}\right) }{\Delta ^{*}\overrightarrow{\Delta }%
^2k_1(k_1-x\Delta )\left( (\overrightarrow{k_1}-x\overrightarrow{\Delta }%
)^2+x(1-x)\overrightarrow{\Delta }^2\right) }
$$
$$
-x^3\overrightarrow{q_1}^2q_1q_2^{*}\,\frac{x\Delta ^{*}k_1\left(
(x-3)\Delta +k_1\right) +\Delta ^2\left( x(2-x)\Delta
^{*}+(1-x)^2k_1^{*}\right) }{\Delta ^{*}\overrightarrow{\Delta }%
^2k_1(k_1-\Delta )\left( (\overrightarrow{k_1}-x\overrightarrow{\Delta }%
)^2+x(1-x)\overrightarrow{\Delta }^2\right) }
$$
$$
+x(1-x)^2\overrightarrow{q_1}^2q_1^{*}q_2\,\frac{x\Delta ^{*}k_1\left(
(x-3)\Delta +k_1\right) +\Delta ^2\left( x(2-x)\Delta
^{*}+(1-x)^2k_1^{*}\right) }{\Delta \overrightarrow{\Delta }%
^2k_1(k_1^{*}-x\Delta ^{*})\left( (\overrightarrow{k_1}-x\overrightarrow{%
\Delta })^2+x(1-x)\overrightarrow{\Delta }^2\right) }
$$
$$
+(1-x)^3q_1q_2^{*}\,\frac{(1-x)q_1^{*}(\Delta -k_1)\left( \Delta
-(2+x)q_1-k_1\right) +q_1^2\left( q_1^{*}-x^2(q_2^{*}+k_1^{*})\right) }{%
\Delta ^{*}(\Delta -k_1)(k_1-x\Delta )\left( (\overrightarrow{\Delta }-
\overrightarrow{k_1}-(1-x)\overrightarrow{q_1})^2+x(1-x)\overrightarrow{q_1}%
^2\right) }
$$
$$
-(1-x)^3q_1q_2^{*}\,\frac{(1-x)q_1^{*}(\Delta -k_1)\left( \Delta
-(2+x)q_1-k_1\right) +q_1^2\left( q_1^{*}-x^2(q_2^{*}+k_1^{*})\right) }{%
\Delta ^{*}(\Delta -k_1)k_1\left( (\overrightarrow{\Delta }-\overrightarrow{%
k_1}-(1-x)\overrightarrow{q_1})^2+x(1-x)\overrightarrow{q_1}^2\right) }
$$
$$
+x^2(1-x)q_1^{*}q_2\,\frac{(1-x)q_1^{*}(\Delta -k_1)\left( \Delta
-(2+x)q_1-k_1\right) +q_1^2\left( q_1^{*}-x^2(q_2^{*}+k_1^{*})\right) }{%
\Delta (\Delta -k_1)(k_1^{*}-x\Delta ^{*})\left( (\overrightarrow{\Delta }-
\overrightarrow{k_1}-(1-x)\overrightarrow{q_1})^2+x(1-x)\overrightarrow{q_1}%
^2\right) }
$$
$$
+(1-x)^3\overrightarrow{q_1}^2q_1q_2^{*}\,\frac{(1-x)\Delta ^{*}(\Delta
-k_1)\left( (1+x)\Delta +k_1\right) -\Delta ^2\left( \Delta
^{*}-x^2k_1^{*}\right) }{\Delta ^{*}\overrightarrow{\Delta }^2(k_1-x\Delta
)(\Delta -k_1)\left( (\overrightarrow{k_1}-x\overrightarrow{\Delta }%
)^2+x(1-x)\overrightarrow{\Delta }^2\right) }
$$
$$
-(1-x)^3\overrightarrow{q_1}^2q_1q_2^{*}\,\frac{(1-x)\Delta ^{*}(\Delta
-k_1)\left( (1+x)\Delta +k_1\right) -\Delta ^2\left( \Delta
^{*}-x^2k_1^{*}\right) }{\Delta ^{*}\overrightarrow{\Delta }^2k_1(\Delta
-k_1)\left( (\overrightarrow{k_1}-x\overrightarrow{\Delta })^2+x(1-x)
\overrightarrow{\Delta }^2\right) }
$$
$$
+x^2(1-x)\overrightarrow{q_1}^2q_1^{*}q_2\,\frac{(1-x)\Delta ^{*}(\Delta
-k_1)\left( (1+x)\Delta +k_1\right) -\Delta ^2\left( \Delta
^{*}-x^2k_1^{*}\right) }{\Delta \overrightarrow{\Delta }^2(k_1^{*}-x\Delta
^{*})(\Delta -k_1)\left( (\overrightarrow{k_1}-x\overrightarrow{\Delta }%
)^2+x(1-x)\overrightarrow{\Delta }^2\right) }
$$
$$
+\frac{(q_1q_2^{*})^2}{(\Delta ^{*})^2}\left( \frac{x^2}{k_1}-\frac{x^2}{%
k_1-x\Delta }\right) \left( \frac{(1-x)^2}{k_1-x\Delta }-\frac{(1-x)^2}{%
k_1-\Delta }\right) -\frac{x^2(1-x)^2(q_1^{*}q_2)^2}{\Delta
^2(k_1^{*}-x\Delta ^{*})^2}
$$
$$
+\frac{\overrightarrow{q_1}^2\overrightarrow{q_2}^2}{\overrightarrow{\Delta
}%
^2}\left( \frac{(1-x)^4}{k_1(k_1^{*}-x\Delta ^{*})}+\frac{x^4}{(k_1-\Delta
)(k_1^{*}-x\Delta ^{*})}-\frac{x^4+(1-x)^2}{(\overrightarrow{k_1}-x
\overrightarrow{\Delta })^2}\right) .
$$

Let us consider now the quark-anti-quark production. The squared amplitude $%
c(k_1,k_2)$ is given below:

$$
\frac 1{x^2}\mid c(k_1,k_2)\mid ^2=
$$
$$
\frac{\mid (1-x)q_1k_1^{*}-xq_1^{*}k_1+x\overrightarrow{q_1}^2\mid ^2}{%
\left( (\overrightarrow{k_1}-x\overrightarrow{q_1})^2+x(1-x)\overrightarrow{%
q_1}^2\right) ^2}+\frac{\overrightarrow{q_1}^4}{\overrightarrow{\Delta }^4}%
\, \frac{\mid (1-x)\Delta k_1^{*}-x\Delta ^{*}k_1+x\overrightarrow{\Delta }%
^2\mid ^2}{\left( (\overrightarrow{k_1}-x\overrightarrow{\Delta })^2+x(1-x)
\overrightarrow{\Delta }^2\right) ^2}
$$
$$
+2Re\,\left( -\,\frac{\overrightarrow{q_1}^2}{\overrightarrow{\Delta }^2}\,
\frac{\left( (1-x)q_1k_1^{*}-xq_1^{*}k_1+x\overrightarrow{q_1}^2\right)
\left( (1-x)\Delta ^{*}k_1-x\Delta k_1^{*}+x\overrightarrow{\Delta }%
^2\right) }{\left( (\overrightarrow{k_1}-x\overrightarrow{q_1})^2+x(1-x)
\overrightarrow{q_1}^2\right) \left( (\overrightarrow{k_1}-x\overrightarrow{%
\Delta })^2+x(1-x)\overrightarrow{\Delta }^2\right) }\right.
$$
$$
\left. +\frac{(1-x)q_1^{*}k_1-xq_1k_1^{*}+x\overrightarrow{q_1}^2}{(
\overrightarrow{k_1}-x\overrightarrow{q_1})^2+x(1-x)\overrightarrow{q_1}^2}%
\left( \frac{(1-x)q_1q_2^{*}}{\Delta ^{*}(k_1-x\Delta )}-\frac{xq_1^{*}q_2}{%
\Delta (k_1^{*}-x\Delta ^{*})}\right) \right.
$$
$$
\left. -\frac{\overrightarrow{q_1}^2}{\overrightarrow{\Delta }^2}\,\frac{%
(1-x)\Delta ^{*}k_1-x\Delta k_1^{*}+x\overrightarrow{\Delta }^2}{(
\overrightarrow{k_1}-x\overrightarrow{\Delta })^2+x(1-x)\overrightarrow{%
\Delta }^2}\left( \frac{(1-x)q_1q_2^{*}}{\Delta ^{*}(k_1-x\Delta )}-\frac{%
xq_1^{*}q_2}{\Delta (k_1^{*}-x\Delta ^{*})}\right) \right.
$$
$$
\left. -\frac{x(1-x)q_1^2(q_2^{*})^2}{(\Delta ^{*})^2(k_1-x\Delta )^2}%
\,\right) \,+\frac{\overrightarrow{q_1}^2\overrightarrow{q_2}^2\left(
x^2+(1-x)^2\right) }{\overrightarrow{\Delta }^2(\overrightarrow{k_1}-x
\overrightarrow{\Delta })^2}\,.
$$

This expression can be written in the following form convenient for the
subsequent integration:%
$$
\frac 1{x^2}\,\mid c(k_1,k_2)\mid ^2=
$$
$$
x(1-x)\left( \frac{\overrightarrow{q_1}^4\left( 6x(1-x)-1\right) +4(1-2x)
\overrightarrow{q_1}^2\overrightarrow{q_1}(\overrightarrow{k_1}-x
\overrightarrow{q_1})-2Re\,q_1^2(k_1^{*}-xq_1^{*})^2}{\left( (
\overrightarrow{k_1}-x\overrightarrow{q_1})^2+x(1-x)\overrightarrow{q_1}%
^2\right) ^2}\right.
$$
$$
\left. +\frac{\overrightarrow{q_1}^4}{\overrightarrow{\Delta }^4}\,\frac{
\overrightarrow{\Delta }^4\left( 6x(1-x)-1\right) +4(1-2x)\overrightarrow{%
\Delta }^2\overrightarrow{\Delta }(\overrightarrow{k_1}-x\overrightarrow{%
\Delta })-2Re\,\Delta ^2(k_1^{*}-x\Delta ^{*})^2}{\left(
(\overrightarrow{k_1%
}-x\overrightarrow{\Delta })^2+x(1-x)\overrightarrow{\Delta }^2\right) ^2}%
\right)
$$
$$
+2\frac{\overrightarrow{q_1}^2}{\overrightarrow{\Delta }^2}\,Re\,\frac{%
x(1-x)(2k_1^2 q_1^{*}\Delta ^{*}-k_1 \overrightarrow{q_1}^2 \Delta^* -q_1
\overrightarrow{\Delta}^2 k_1^{*})+x^2\Delta q_1^{*}\left( k_1\Delta
^{*}+q_1k_1^{*}-q_1\Delta ^{*}\right) }{\left( (\overrightarrow{k_1}-x
\overrightarrow{q_1})^2+x(1-x) \overrightarrow{q_1}^2\right) \left( (
\overrightarrow{k_1}-x\overrightarrow{\Delta })^2+x(1-x)\overrightarrow{%
\Delta }^2\right) }
$$
$$
+\frac{\overrightarrow{q_1}^2}{\overrightarrow{\Delta }^2}\left(
x^2+(1-x)^2\right) \,\left( \frac{-2x\overrightarrow{q_1}\overrightarrow{%
\Delta }(2\overrightarrow{k_1}\overrightarrow{q_1}-\overrightarrow{q_1}^2)
}{%
\left( (\overrightarrow{k_1}-x\overrightarrow{q_1})^2+x(1-x)
\overrightarrow{%
q_1}^2\right) \left( (\overrightarrow{k_1}-x\overrightarrow{\Delta }%
)^2+x(1-x)\overrightarrow{\Delta }^2\right) }\right.
$$
$$
\left. +\frac{\overrightarrow{q_1}^2-\overrightarrow{q_2}^2}{(
\overrightarrow{k_1}-x\overrightarrow{q_1})^2+x(1-x)\overrightarrow{q_1}^2}-
\frac{\overrightarrow{q_1}^2}{(\overrightarrow{k_1}-x\overrightarrow{\Delta
}%
)^2+x(1-x)\overrightarrow{\Delta }^2}+\frac{\overrightarrow{q_2}^2}{(
\overrightarrow{k_1}-x\overrightarrow{\Delta })^2}\right)
$$
$$
+2Re\,\left( \frac{xq_1q_2^{*}}{\Delta ^{*}(k_1-x\Delta )}\,\frac{%
(1-x)q_1(q_1^{*}-2k_1^{*})+\Delta q_1^{*}(x^2+(1-x)^2)-x\overrightarrow{q_1}%
^2}{(\overrightarrow{k_1}-x\overrightarrow{q_1})^2+x(1-
x)\overrightarrow{q_1}%
^2}\right.
$$
$$
\left. +\frac{\overrightarrow{q_1}^2}{\overrightarrow{\Delta }^2}\,\,\frac{%
2\Delta k_1^{*}-2(1-x)\overrightarrow{\Delta }^2}{(\overrightarrow{k_1}-
\overrightarrow{x\Delta })^2+x(1-x)\overrightarrow{\Delta }^2}\,\,\frac{%
x(1-x)q_1q_2^{*}}{\Delta ^{*}(k_1-x\Delta )}-\frac{x(1-x)(q_1q_2^{*})^2}{%
(\Delta ^{*})^2(k_1-x\Delta )^2}\right) \,.
$$

\thinspace The interference term for the quark-anti-quark production is
given below:%
$$
\frac 1{x(1-x)}\,c(k_1,k_2)\,c(k_2,k_1)=
$$
$$
\frac{\overrightarrow{k_1}^2\overrightarrow{q_1}^2\left( x^2+(1-x)^2\right)
-x(1-x)\,2Re\,(k_1^2q_1^{*}\,^2)+x\overrightarrow{q_1}^2\left( xq_1\Delta
^{*}+(1-x)q_1^{*}q_2\right) }{\left( (\overrightarrow{k_1}-x\overrightarrow{%
q_1})^2+x(1-x)\overrightarrow{q_1}^2\right) \left( (\overrightarrow{k_1}+
\overrightarrow{q_2}-x\overrightarrow{q_1})^2+x(1-x)\overrightarrow{q_1}%
^2\right) }
$$
$$
+\frac{xq_1^{*}k_1\left( (1-x)q_1^{*}\Delta -xq_1\Delta ^{*}\right)
+k_1^{*}\left( (1-2x)\overrightarrow{q_1}^2q_1+(1-x)q_1(xq_1\Delta
^{*}-(1-x)q_1^{*}\Delta )\right) }{\left( (\overrightarrow{k_1}-x
\overrightarrow{q_1})^2+x(1-x)\overrightarrow{q_1}^2\right) \left( (
\overrightarrow{k_1}+\overrightarrow{q_2}-x\overrightarrow{q_1})^2+x(1-x)
\overrightarrow{q_1}^2\right) }
$$
$$
+\frac{\overrightarrow{q_1}^4}{\overrightarrow{\Delta }^4}\,\frac{
\overrightarrow{k_1}^2\overrightarrow{\Delta }^2\left( x^2+(1-x)^2\right)
-x(1-x)\,2Re\,(k_1^2\Delta ^{*}{}^2)+x(1-2x)\overrightarrow{\Delta }%
^2\,2Re\,(k_1\Delta ^{*})+x^2\overrightarrow{\Delta }^4}{\left( (
\overrightarrow{k_1}-x\overrightarrow{\Delta })^2+x(1-x)\overrightarrow{%
\Delta }^2\right) ^2}
$$
$$
-\frac{\overrightarrow{q_1}^2}{\overrightarrow{\Delta }^2}\,\frac{
\overrightarrow{k_1}^2\left( (1-x)^2\Delta q_1^{*}+x^2\Delta ^{*}q_1\right)
-x(1-x)\,2Re\,\left( k_1^2\Delta ^{*}q_1^{*}\right) +x\overrightarrow{\Delta
}^2\left( xq_1\Delta ^{*}+(1-x)q_1^{*}q_2\right) }{\left( (\overrightarrow{%
k_1}+\overrightarrow{q_2}-x\overrightarrow{q_1})^2+x(1-
x)\overrightarrow{q_1}%
^2\right) \left( (\overrightarrow{k_1}-x\overrightarrow{\Delta })^2+x(1-x)
\overrightarrow{\Delta }^2\right) }
$$
$$
-\frac{\overrightarrow{q_1}^2}{\overrightarrow{\Delta }^2}\,\frac{x\Delta
^{*}k_1\left( (1-x)q_1^{*}(2\Delta -q_1)-xq_1\Delta ^{*}\right) +\Delta
k_1^{*}\left( (1-x)(xq_1\Delta ^{*}+(1-x)q_1^{*}q_2)-x^2q_1\Delta
^{*}\right) }{\left( (\overrightarrow{k_1}+\overrightarrow{q_2}-x
\overrightarrow{q_1})^2+x(1-x)\overrightarrow{q_1}^2\right) \left( (
\overrightarrow{k_1}-x \overrightarrow{\Delta })^2+x(1-x)\overrightarrow{%
\Delta }^2\right) }
$$
$$
-\frac{\overrightarrow{q_1}^2}{\overrightarrow{\Delta }^2}\,\frac{
\overrightarrow{k_1}^2\left( x^2q_1^{*}\Delta +(1-x)^2q_1\Delta ^{*}\right)
-x(1-x)\,2Re\,\left( k_1^2q_1^{*}\Delta ^{*}\right) +x^2\overrightarrow{q_1}%
^2\overrightarrow{\Delta }^2}{\left( (\overrightarrow{k_1}-x\overrightarrow{%
q_1})^2+x(1-x)\overrightarrow{q_1}^2\right) \left( (\overrightarrow{k_1}-x
\overrightarrow{\Delta })^2+x(1-x)\overrightarrow{\Delta }^2\right) }
$$
$$
-\frac{\overrightarrow{q_1}^2}{\overrightarrow{\Delta }^2}\,\frac{%
xq_1^{*}\Delta ^{*}k_1\left( (1-x)q_1-x\Delta \right) +xq_1\Delta
k_1^{*}\left( (1-x)\Delta ^{*}-xq_1^{*}\right) } {\left(
(\overrightarrow{k_1%
}-x\overrightarrow{q_1})^2+x(1-x)\overrightarrow{q_1}^2\right) \left( (
\overrightarrow{k_1}-x\overrightarrow{\Delta })^2+x(1-x)\overrightarrow{%
\Delta }^2\right) }
$$
$$
+\frac{x^2\overrightarrow{q_1}^2q_2^{*}k_1-xq_1q_2^{*}\left(
(1-x)q_1k_1^{*}+x\overrightarrow{q_1}^2\right) }{\Delta ^{*}(k_1-x\Delta
)\left( (\overrightarrow{k_1}-x\overrightarrow{q_1})^2+x(1-
x)\overrightarrow{%
q_1}^2\right) }
$$
$$
+\frac{(1-x)^2\overrightarrow{q_1}^2q_2k_1^{*}-x(1-x)q_1^{*2}q_2(k_1-q_1)}{%
\Delta (k_1^{*}-x\Delta ^{*})\left( (\overrightarrow{k_1}-x\overrightarrow{%
q_1})^2+x(1-x)\overrightarrow{q_1}^2\right) }
$$
$$
+\frac{(1-x)^2\overrightarrow{q_1}^2q_2^{*}k_1+(1-x)q_1q_2^{*}\left(
xq_1(\Delta ^{*}-k_1^{*})+(1-x)q_1^{*}q_2\right) }{\Delta ^{*}(k_1-x\Delta
)\left( (\overrightarrow{k_1}+\overrightarrow{q_2}-x\overrightarrow{q_1}%
)^2+x(1-x)\overrightarrow{q_1}^2\right) }
$$
$$
+\frac{x^2\overrightarrow{q_1}^2q_2k_1^{*}-xq_1^{*}q_2\left( xq_1\Delta
^{*}+(1-x)q_1^{*}(k_1+q_2)\right) }{\Delta (k_1^{*}-x\Delta ^{*})\left( (
\overrightarrow{k_1}+\overrightarrow{q_2}-x\overrightarrow{q_1})^2+x(1-x)
\overrightarrow{q_1}^2\right) }
$$
$$
-\frac{\overrightarrow{q_1}^2q_1q_2^{*}}{\overrightarrow{\Delta }^2\Delta
^{*}}\,\frac{\Delta ^{*}k_1(x^2+(1-x)^2)-x\Delta \left(
2(1-x)k_1^{*}-(1-2x)\Delta ^{*}\right) }{(k_1-x\Delta )\left( (
\overrightarrow{k_1}-x\overrightarrow{\Delta })^2+x(1-x)\overrightarrow{%
\Delta }^2\right) }
$$
$$
-\frac{\overrightarrow{q_1}^2q_1^{*}q_2}{\overrightarrow{\Delta }^2\Delta }%
\, \frac{\Delta k_1^{*}((1-x)^2+x^2)-2x(1-x)\Delta ^{*}k_1+x(1-2x)
\overrightarrow{\Delta }^2}{(k_1^{*}-x\Delta ^{*})\left(
(\overrightarrow{k_1%
}-x\overrightarrow{\Delta })^2+x(1-x)\overrightarrow{\Delta }^2\right) }
$$
$$
+\left( x^2+(1-x)^2\right) \frac{\overrightarrow{q_1}^2\overrightarrow{q_2}%
^2 }{\overrightarrow{\Delta }^2(\overrightarrow{k_1}-x\overrightarrow{\Delta
})^2}-2x(1-x)\,Re\,\frac{q_1^2q_2^{*2}}{\Delta ^{*2}(k_1-x\Delta )^2}
$$

\def\beq#1{\begin{equation}\label{#1}}
\def\beeq#1{\begin{eqnarray}\label{#1}}
\def\eeq{\end{equation}}
\def\eeeq{\end{eqnarray}}
\appendix
%\chapter{}
%\input{appendixd}
%\chapter{}
%\input{appendixf}
%%
%%
\newpage
\begin{figure}[t]
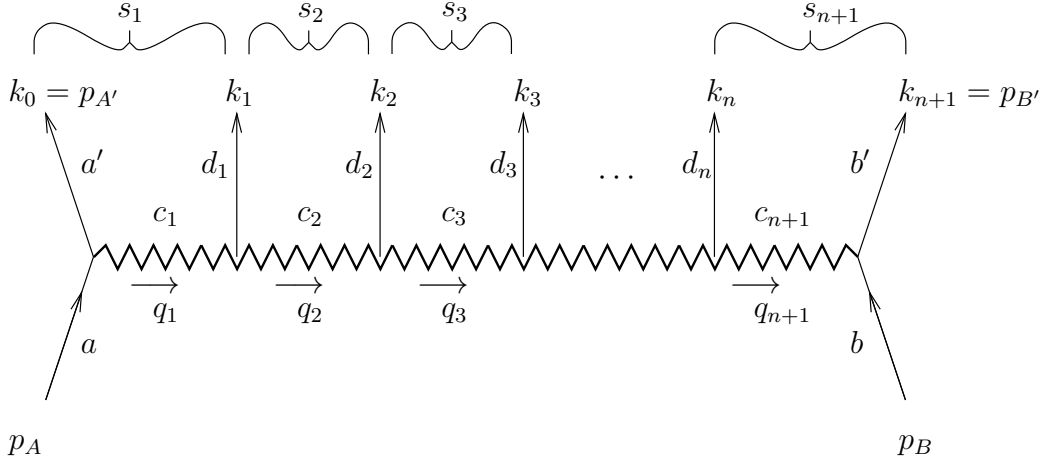

\input fig1.pstex_t
\caption{
Gluon production amplitude in the multi-Regge kinematics;
the crossing channel gluons with momenta $q_1,q_2,...,q_{n+1}$
are reggeized.}
\end{figure}
\begin{figure}[b]
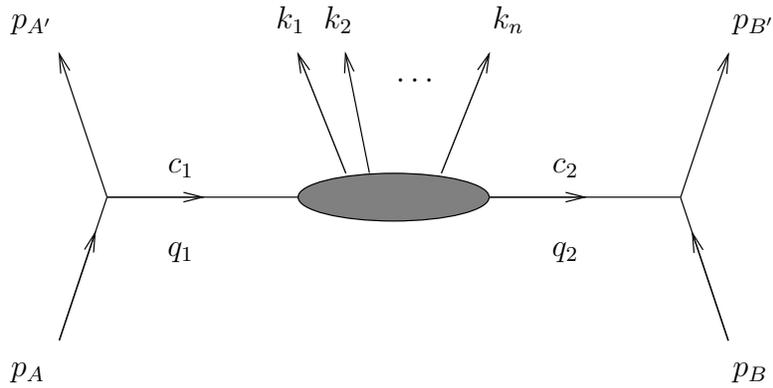

\begin{center}
\input fig2.pstex_t
\caption{Production amplitude in the quasi-multi-Regge
kinematics; the invariant mass $\protect\sqrt{\kappa}$ for the produced
particles in the central rapidity region is fixed.}
\end{center}
\end{figure}
\newpage
\begin{figure}[t]
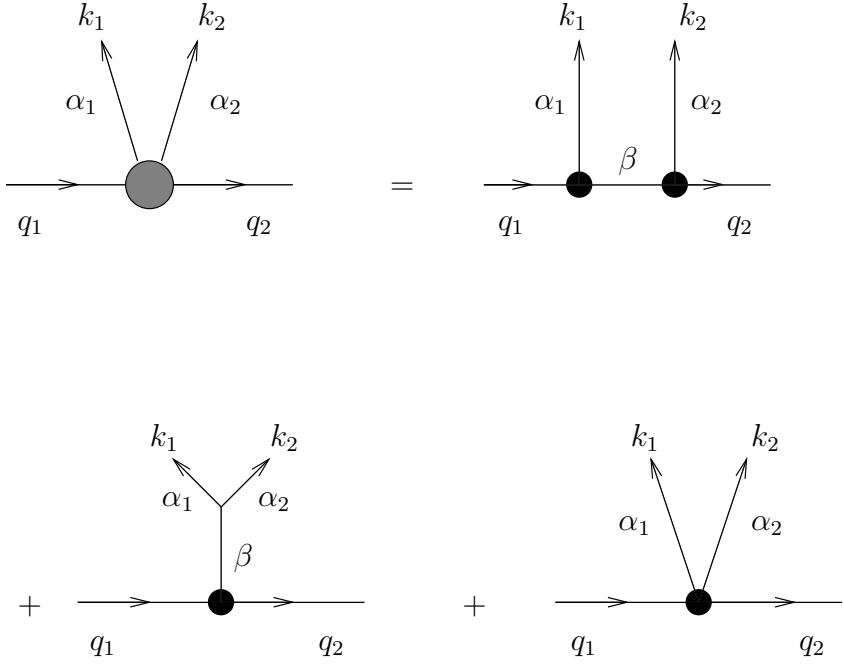

\input fig3.pstex_t
\caption{
Two gluon production amplitude $A^{\alpha_1,\alpha_2}(k_1,k_2)$
expressed in terms of the diagrams with the usual and effective vertices.}
\end{figure}
\begin{figure}[b]
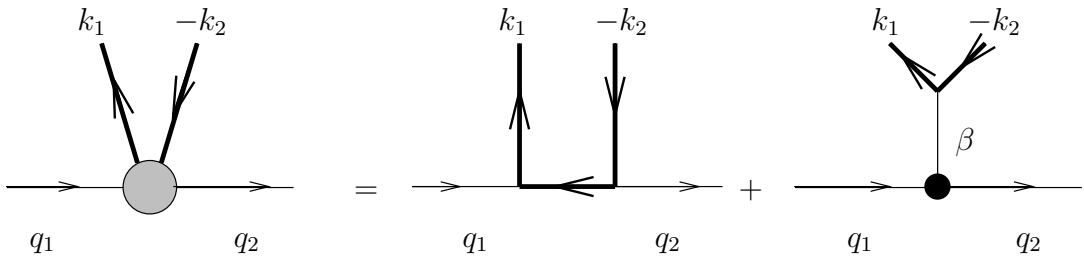

\input fig4.pstex_t
\caption{
Quark-antiquark production amplitude $b(k_1,k_2)$ expressed in terms
of the diagrams with the usual and effective vertices.}
\end{figure}
\end{document}